\documentclass{PoS}

\usepackage{inputenc}
\usepackage{unites2e}
 \usepackage{warn}
 \usepackage{wrapfig}
\usepackage{bibspacing}
\def\grays{$\gamma$-rays}
\def\gray{$\gamma$-ray}

\newcommand{\HESS}{H.E.S.S.}
\def\LSI{LS~I~+61$^\circ$~303}
\def\LAT{{\it Fermi}-LAT}

\title{The Very High Energy Sky from $\mathbf{\sim 20}$ GeV to Hundreds of TeV - Selected Highlights}

\ShortTitle{The Very High Energy Sky from $\sim 20$ GeV to Hundreds of TeV - Selected Highlights}

\author{\speaker{Mathieu de Naurois}\\
        LLR - IN2P3/CNRS - Ecole Polytechnique\\
        E-mail: \email{denauroi@in2p3.fr}}

\abstract{After nearly a decade of operation, the three major arrays of atmospheric Cherenkov telescopes have 
revolutionized our view of the Very High Energy Universe, unveiling more than 100 sources of various types. 
MAGIC, consisting of two $17\U{m}$ diameter telescopes on the Canary island of La Palma, and VERITAS, with four $12\U{m}$ telescopes 
installed in southern Arizona, USA, have primarily explored the extragalactic sky, where the majority of the sources are active galactic nuclei (AGN), 
with \gray\ emission originating in their relativistic jets. 
In July 2014 MAGIC discovered the most distant source of very high energy $\gamma$ rays, 
the gravitationally lensed blazar S3\  0218 residing at the redshift of $z=0.944$. 
The flat spectrum radio quasar PKS~1441+25 ($z=0.939$), observed by MAGIC \& VERITAS in 2015, showed a strong flaring activity over a time span of several weeks. 
Rapid variability from various BL Lacertae objects, down to minute timescales, has been observed by the three experiments, 
and measurement of their high-energy spectra allows the level of extragalactic background light to be constrained.

Since the commissioning of the fifth, large ($28\U{m}$ diameter) telescope in December 2012, \HESS-II is the only 
array operating telescopes of different sizes together (``{\it hybrid array }''). The largest Cerenkov telescope
ever built, CT5, provides an energy threshold of $\sim 20\U{GeV}$.
With its broad energy range, \HESS\ explored 
the Galactic Plane with unprecedented sensitivity. The legacy release of the \HESS\ Galactic Plane Survey, 
consisting of 2800 hours of observations of the Galactic disk, reveals major new results. This is the first high-resolution ($\sim 0.1\U{deg}$)  
sensitive ($\sim 2\%$ Crab Nebula point-source sensitivity) survey of the Milky Way in VHE $\gamma$ rays.

The Milky Way harbors a large variety of high energy sources of various types. In recent years, deep observations of several key 
Galactic regions of utmost importance for this field have been conducted by the three experiments. Among them are the Galactic Center 
region and its halo (particularly relevant for dark matter searches), the Cygnus region and its mysterious Milagro sources, the Crab Nebula 
and pulsar (surprisingly showing pulsed emission up to above 1 TeV), the iconic \gray\ supernova remnant RX\ J1713.7-3946 and other SNRs 
such as Tycho,  the Vela pulsar and several binary systems such as LS~5039, PSR~B1259-63, HESS~J0632+057 and \LSI. 
Joint observations on several of 
these objects proved to be the most efficient way to understand their nature. A deep observation of the Large Magellanic Cloud revealed, for the first time, 
spectacular and powerful accelerators of stellar origin outside our own galaxy.

Highlights of these observations with \HESS, MAGIC and VERITAS have been presented and discussed at the conference.
}

\FullConference{The 34th International Cosmic Ray Conference,\\
		30 July- 6 August, 2015\\
		The Hague, The Netherlands}

\begin{document}

\section*{Introduction}

In the last years, the field of very high energy (VHE) \gray astronomy evolved in several directions,
taking advantage of the improved sensitivities over time (see Fig.~\ref{fig:SensVsTime}):

\begin{figure}[b!]
\begin{center}
\includegraphics[width=0.7\textwidth]{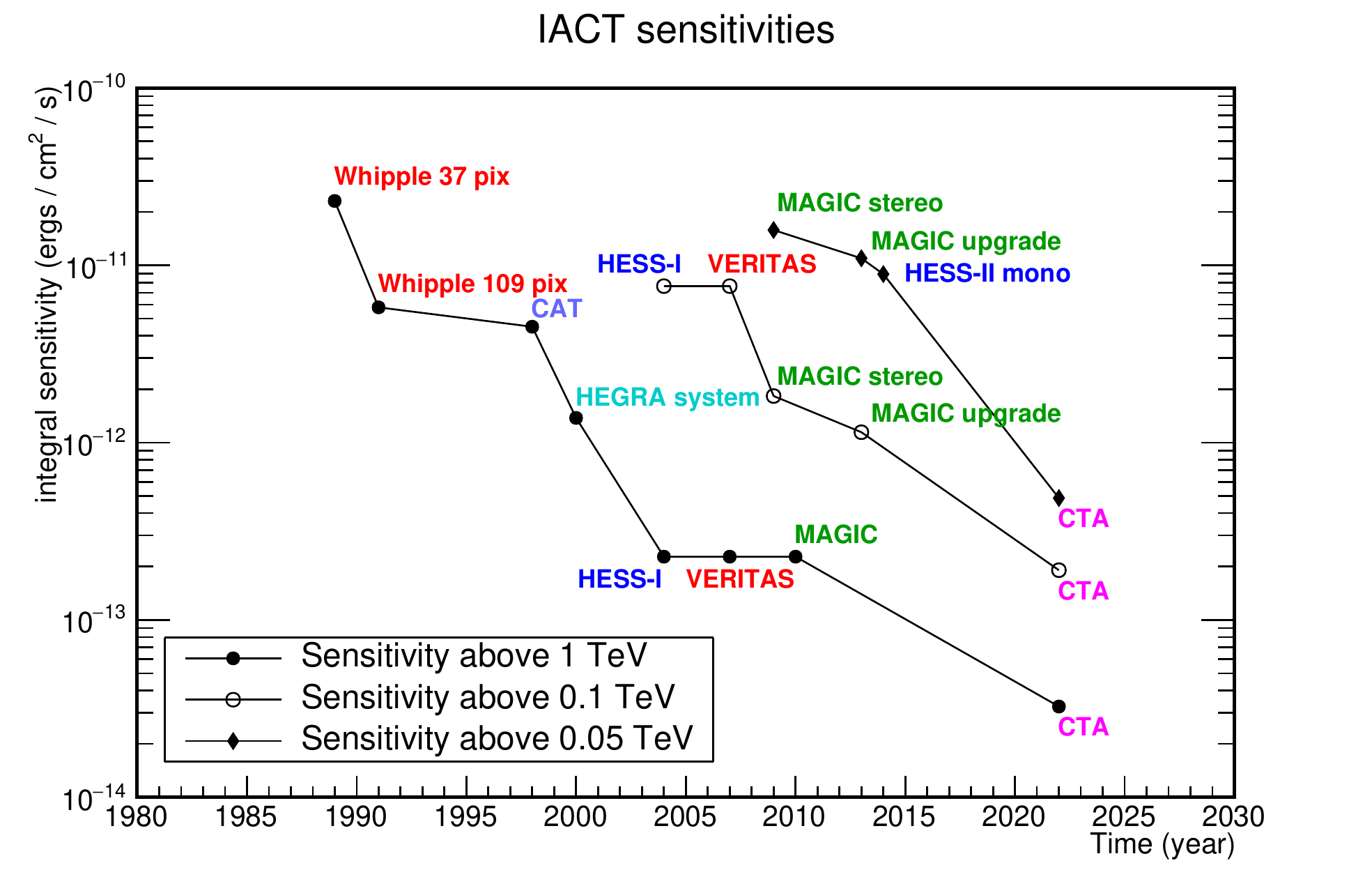} 
\end{center}
\caption{\label{fig:SensVsTime}Evolution of integral sensitivity of the IACTs over time, above 
$50\U{GeV}$ (filled diamonds), $100\U{GeV}$ (open circles) and $1\U{TeV}$ (filled circles)~\cite{CRAS_IACTS}. 
The expected sensitivities of CTA are also shown.}
\end{figure}

\begin{itemize}
\item Large surveys, consisting of several hundreds - or even thousands - of hours have been performed by the \HESS\ experiment in the central regions 
of the Galaxy and by the VERITAS collaboration in the Cygnus region. These surveys provide a unique way to characterize the VHE source population
in a uniform, unbiased way, and open the way to population studies.
\item Deep exposures have been taken on a handful of sources of particular importance (``{\it key science projects}''), allowing in-depth understanding of
the acceleration and radiation mechanisms, but also constraining upper limits on exotic physics and dark matter searches.
This has been made possible by the development of sophisticated analysis techniques and a much deeper understanding of the instruments.
\item A particular effort has been undertaken to reduce the energy threshold of the instruments and bridge the gap to \LAT. 
The VERITAS collaboration~\cite{veritas-highlight-icrc2015} upgraded
its telescopes with high quantum efficiency PMTs in 2012~\cite{veritas-perf-icrc2015}  while MAGIC is since 2009 operating in stereo mode. 
\item The largest Imaging Atmospheric Cerenkov Telescope ever built, namely the fifth telescope of \HESS-II (CT5), has been inaugurated and is now operating regularly,
allowing observations of $\gamma$ rays down to $\sim 10\U{GeV}$.
\HESS\ is currently performing a major upgrade
of the Cerenkov telescopes 1 -- 4~\cite{hess-upgrade-icrc2015} in order to lower their threshold when operated with CT5 and reduce their dead-time.
\item After the success of Milago~\cite{2004ApJ...608..680A,2007ApJ...664L..91A},
the High Altitude Water Cerenkov (HAWC)~\cite{2012APh....35..641A,2013APh....50...26A}, a large water Cerenkov detector with a very large 
field of view, has been inaugurated in March 2015. Very well suited to full-sky surveys, HAWC, has been the subject of a dedicated 
review talk~\cite{hawc-icrc2015} and is therefore not covered by this paper.
\end{itemize}

\section{The \HESS\ Legacy Survey}

\begin{figure}[b!]
\begin{center}
\includegraphics[width=\textwidth]{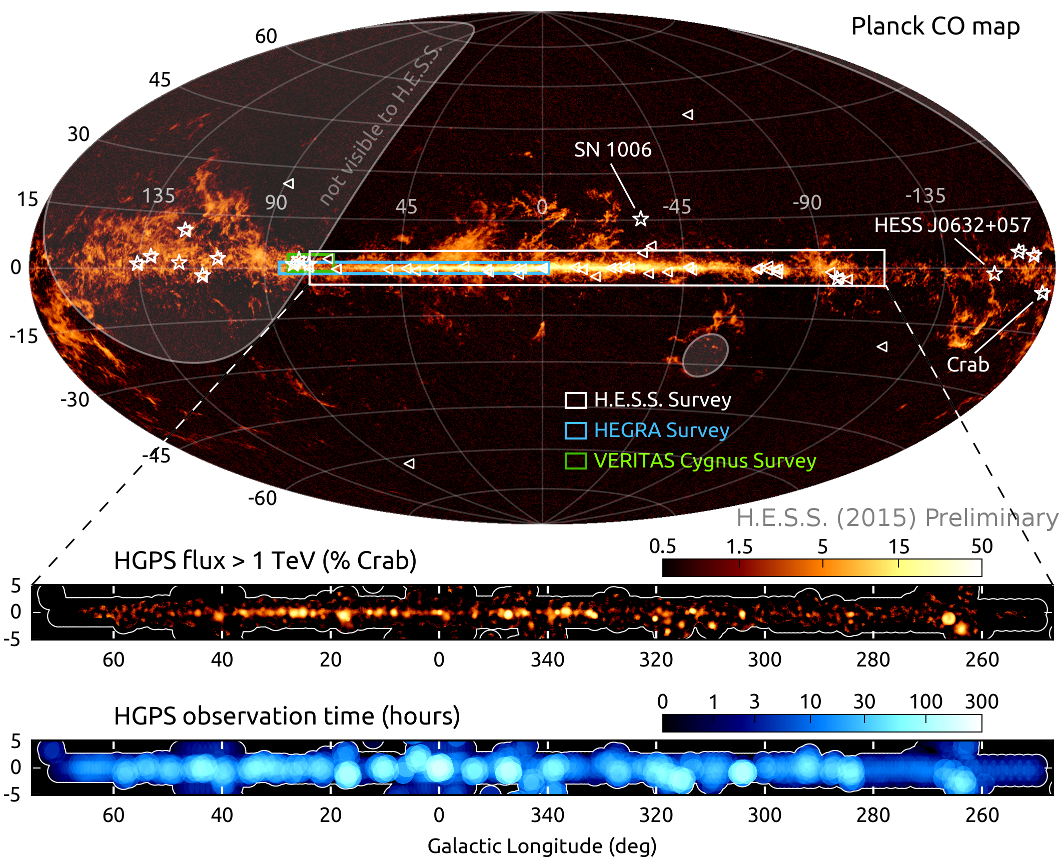}
\end{center}
\caption{\label{fig:HPGS01}\HESS\ Galactic Plane Survey region, flux map and exposure map (from top to bottom). The all-sky image on the top panel
shows a Planck CO Map with \LAT\ identified Galactic 1FHL sources (triangles) and the 15 known Galactic TeV sources (white stars) outside the HGPS region.
The HEGRA Galactic Plane Survey~\cite{hegra-survey} and the VERITAS Cygnus survey~\cite{2009arXiv0912.4492W,veritas-cygnus-icrc2015} regions are illustrated in blue and green, respectively.
From~\cite{hess-hgps-icrc2015}.}
\end{figure}

The \HESS\  Galactic plane survey (HGPS)~\cite{hess-hgps-icrc2015} was performed
with the \HESS-I Cerenkov telescope array in Namibia from 2004 to 2013.
After several publications on a small ($\sim 10\%$) fraction of the current data set~\cite{hess-survey1,hess-survey2},
it is the deepest and most comprehensive, high resolution ($\sim 0.1\U{degrees}$) and sensitive ($\lesssim 2\%$ Crab Nebula point-source sensitivity) 
survey of the Milky Way in very-high-energy $\gamma$ rays\ ($0.2 \lesssim E \lesssim 100\U{TeV}$).
Roughly $\sim 2700$ hours of high-quality observations of the Galactic disk are available in the Galactic longitude
range from 250 to 65 degrees and Galactic latitude $|b| < 3.5\U{degrees}$. The region of the Milky Way covered by the HGPS is depicted
as a white rectangle in Fig.~\ref{fig:HPGS01} and compared to the HEGRA (in blue) and VERITAS Cygnus (in green) surveys.



The HGPS data set combines dedicated survey operations (using a fixed-grid pointing strategy)  with deep observations of sources of particular
interest and follow-up observations of previously discovered sources. Therefore, as shown in Fig.~\ref{fig:HPGS01}, lower panel, the exposure,
and thus the sensitivity, is not uniform across the survey region.
For the first time, an automatic pipeline was used for source extraction, using a likelihood fit of Gaussian components plus diffuse background.
11 complex sources  (e.g. shell supernova remnants) were excluded from the pipeline.
A catalogue of 77 cosmic accelerators  was obtained,
out of which 6 are new sources that were previously unknown or unpublished.
A significant fraction of the VHE population are identified as pulsar wind nebul\ae\ (PWN) or supernova remnants (SNR), 
while the majority remains unidentified or confused.

The HGPS is a unique tool for population studies (e.g. PWN~\cite{hess-pwn-icrc2015} or
SNR~\cite{hess-snrpop-icrc2015} population studies were shown at the conference). The paper and legacy data will be released soon
(Fall 2015), including FITS maps and a source catalog (morphology \& spectra) and will certainly become a reference for the community.

\section{Supernova Remnants as sources of cosmic rays}

Expanding shock waves in SNRs are believed to be able to accelerate cosmic rays up to 
multi-TeV energies through the mechanism of diffusive shock acceleration (DSA), see e.g.~\cite{DruryDSA}.
It was realized very early that, given the rate of $\sim 3$ explosive
events per century,  SNRs are capable of maintaining the galactic 
cosmic ray (CR) flux at the observed level~\cite{Baade1934}, if a fraction of about $10\%$ of their explosion 
energy is converted into cosmic rays. 
About 300 SNRs are currently known in the Milky Way, mostly from their non thermal radio emission~\cite{Green2004,Ferrand2012}. 
Most shell-type SNRs are non-thermal radio emitters, showing evidence
that electrons are accelerated up to at least GeV energies.
For a recent review of diffusive shock acceleration in the 
context of SNRs, see e.g.~\cite{HillasCRReview}.

Several classes of SNRs are potential sites of VHE cosmic-ray acceleration.
Composite SNRs host an energetic pulsar at their center, whose non-thermal outflow
leads to the formation of a PWN, capable of accelerating electrons up to VHE energies.
Young shell-type SNRs are expected to release a significant fraction of their non
thermal output at TeV energies, whereas older SNRs, through interactions with
clouds of interstellar matter, are more likely to shine at GeV energies, thus allowing
the different stages of SNR evolution to be probed using multi-wavelength observations.

In contrast to very high energy electrons, that unavoidably shine in $\gamma$ rays through inverse-Compton (IC) scattering on 
cosmological microwave and infrared backgrounds photons, accelerated protons and ions 
are only revealed when interacting with a dense target, such as a molecular cloud (MC). 
Electrons, on the other hand, suffer from stronger energy losses and have short live times in dense environments.
In the end, the emission mechanism depends more on the environment than on the accelerator itself.

In the last years, SNRs have been firmly established as sources of VHE \gray\ emission.
At least five SNRs with clear shell-type morphology resolved in VHE $\gamma$ rays were detected by \HESS, allowing direct investigation of the SNRs
as sources of cosmic rays. They are all remnants of recent supernovae (less than a few kyr):
RX~J1713.7-3946~\cite{hess-rxj1713-nature,hess-rxj1713-deep}, RX~J0852.04622 (also known as Vela~Junior)~\cite{hess-vela-junior},
SN~1006~\cite{hess-SN1006}, HESS~J1731-347~\cite{hess-1731} and RCW~86~\cite{hess-RCW86}. 
All of them show a very clear correlation between non-thermal X-ray and VHE \gray\ emissions.
Recently, two additional candidates of shell-type SNRs were identified in the HGPS~\cite{hess-snrs-icrc2015}, HESS~J1534-571 and HESS~J1912+101,
the latter being the first TeV-only shell candidate, identified through its morphology and without known counterpart.
Three other shell-type SNRs, Tycho ~\cite{veritas-tycho,veritas-tycho-icrc2015}, Cas~A~\cite{hegra-casA,magic-casA,veritas-casA,veritas-casA-icrc2015} 
and IC~443\cite{magic-ic443,veritas-ic443}
are detected at VHE by MAGIC and VERITAS\footnote{Tycho was first discovered by VERITAS, Cas~A by HEGRA and IC~443 by MAGIC.} and 
at HE~\cite{fermi-tycho,fermi-casA,fermi-ic443-w44}. All three exhibit a rather
soft spectrum, peaking in the GeV range, and consistent with a pion-bump, thus supporting a hadronic scenario.


\subsection{SNR RX~J1713.7-3946}

The SNR RX~J1713.7-3946 is one of the brightest Galactic X-ray SNRs,
possibly associated with the guest star  AD393 which, according to Chinese
astronomers, appeared in the constellation Scorpius~\cite{1997A&A...318L..59W}.
After its discovery by the CANGAROO collaboration~\cite{2002Natur.416..823E},
It was the first young, shell-type SNR to be resolved in VHE $\gamma$ rays~\cite{hess-rxj1713-nature},
with a typical shell morphology consistent with X-ray observations.

There is an ongoing debate in the literature about the origin of the emission.
Recent observations of RX~J1713.7-3946 with the \LAT~\cite{fermi-rxj1713}
still suggest that the dominant emission mechanism might be leptonic,
although a significant contribution from hadronic cosmic rays cannot be excluded. The lack of
X-ray thermal bremsstrahlung emission, predicted in many hadronic scenarios, further support this
interpretation~\cite{Ellison2010}, as well as the close correlation between the X-ray and \gray\ morphologies.
However, \gray\ emission up to $100\U{TeV}$ is difficult to achieve with IC scattering
due to Klein-Nishina suppression of the cross section at high energy. The shape of the high-energy cut off therefore
provides a handle on the acceleration mechanism. 


\begin{figure}[hb!]
\begin{center}
\includegraphics[height=6cm]{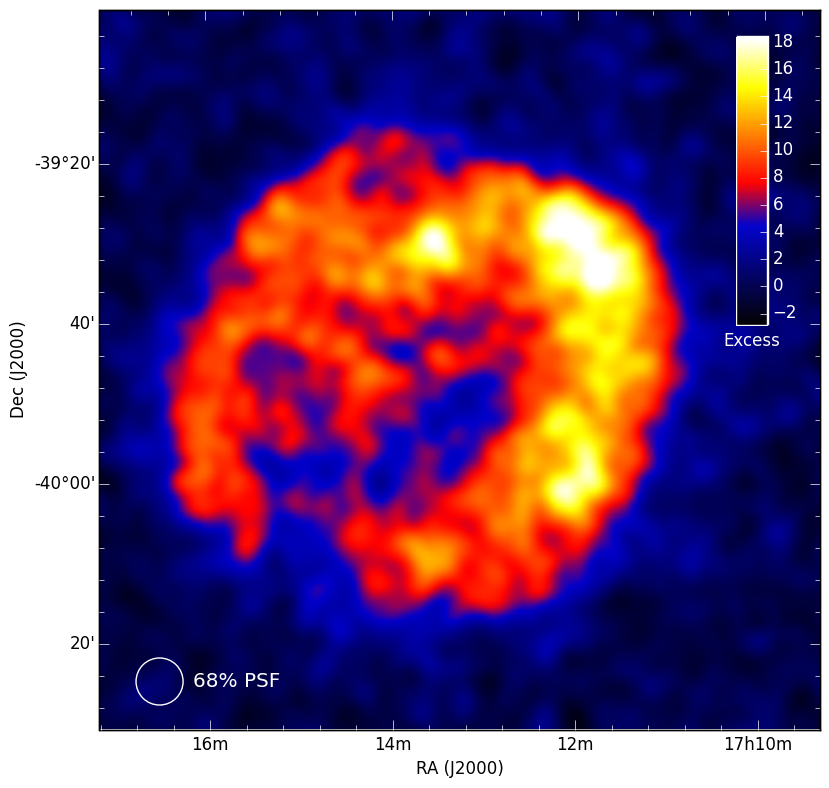}\hfill
\includegraphics[height=6cm]{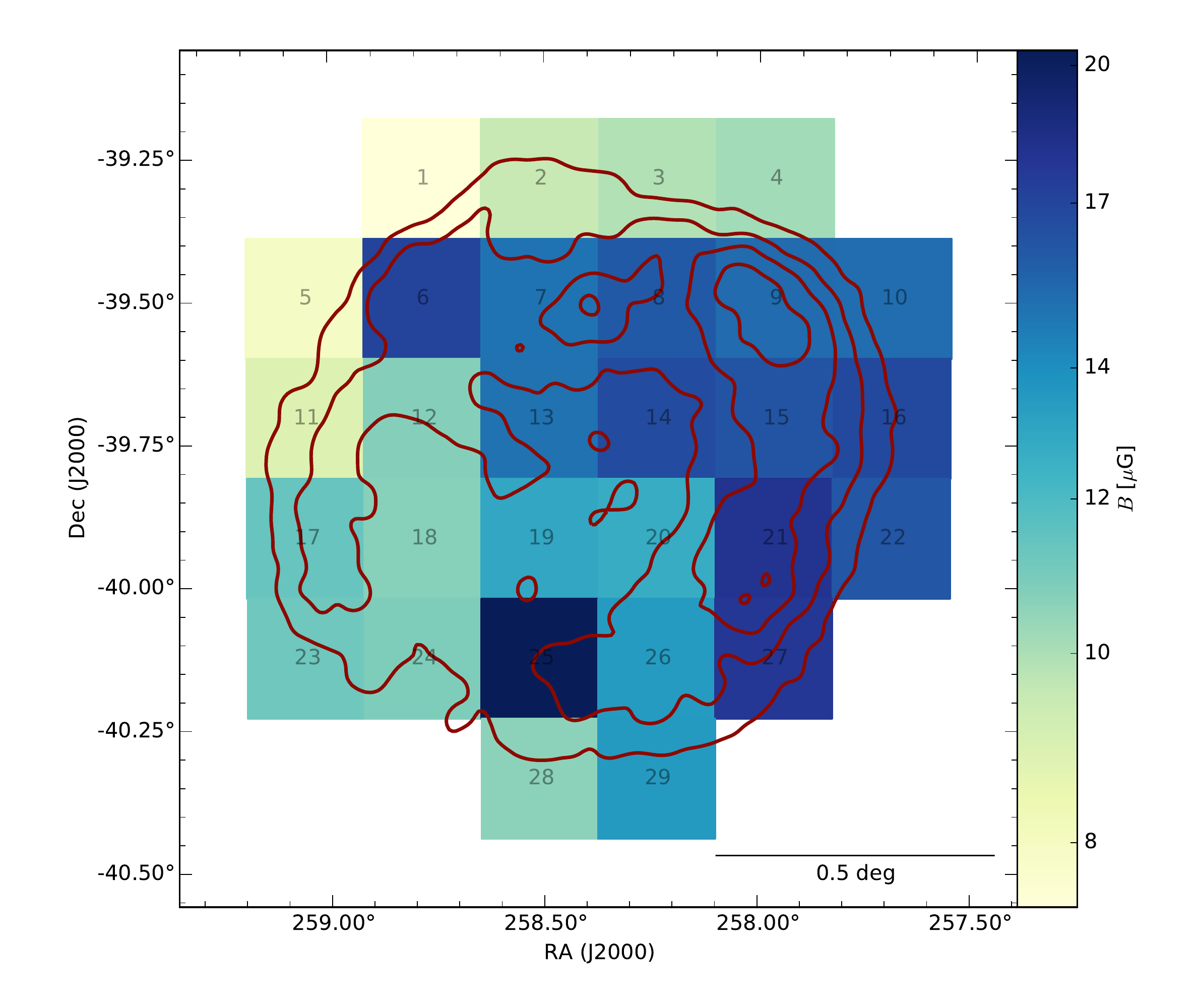}
\end{center}
\caption{\label{fig:RXJ1713}{\bf Left}: \HESS VHE \gray\ image of RX~J1713.7-3946. {\bf Right}: Map of magnetic field under a leptonic scenario~\cite{hess-rxj1713-icrc2015}.}
\end{figure}

Under a purely IC scenario for the GeV-TeV spectrum, the model-independent derived electron spectrum exhibits a break at $\sim 2\U{TeV}$,
which could only be obtained with a magnetic field of $> 100\U{\mathrm{\mu}G}$, in conflicts with direct estimations.
Alternately, a very high photon field energy density ($\sim 140 \U{eV}\UU{cm}{-3}$)
could provide the required target for strong IC emission, but this value exceeds by far recent estimations of the target photon field.
In summary, it appears very difficult, given the observational constrains, to explain the \LAT\ and \HESS\ data in a consistent manner, with a purely leptonic
model based on a single population of electrons.



\begin{figure}[th!]
\begin{center}
\includegraphics[height=6cm]{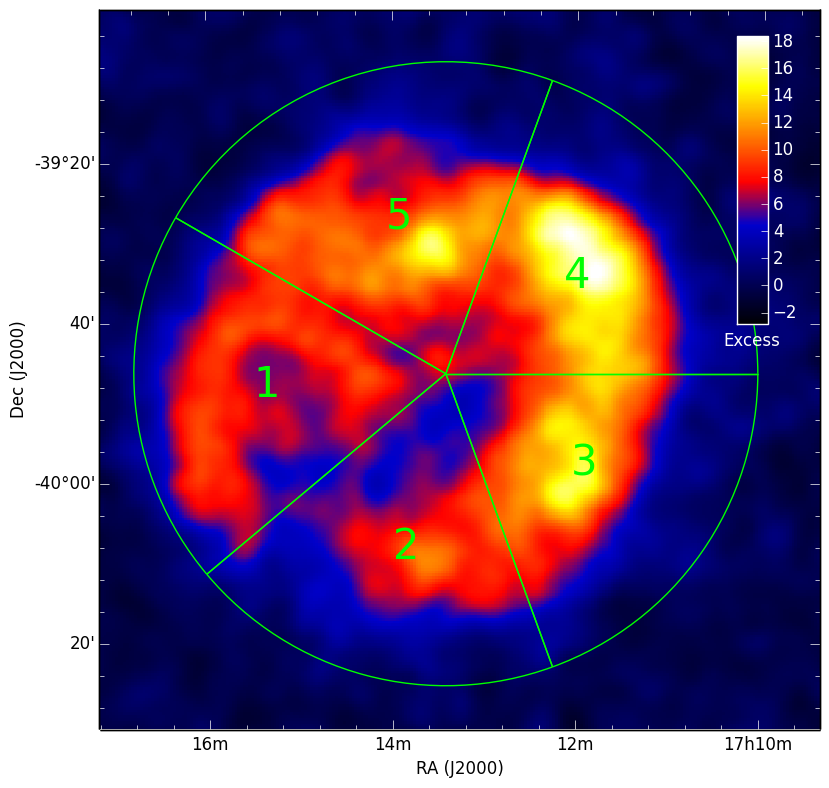}\hfill
\includegraphics[height=6cm]{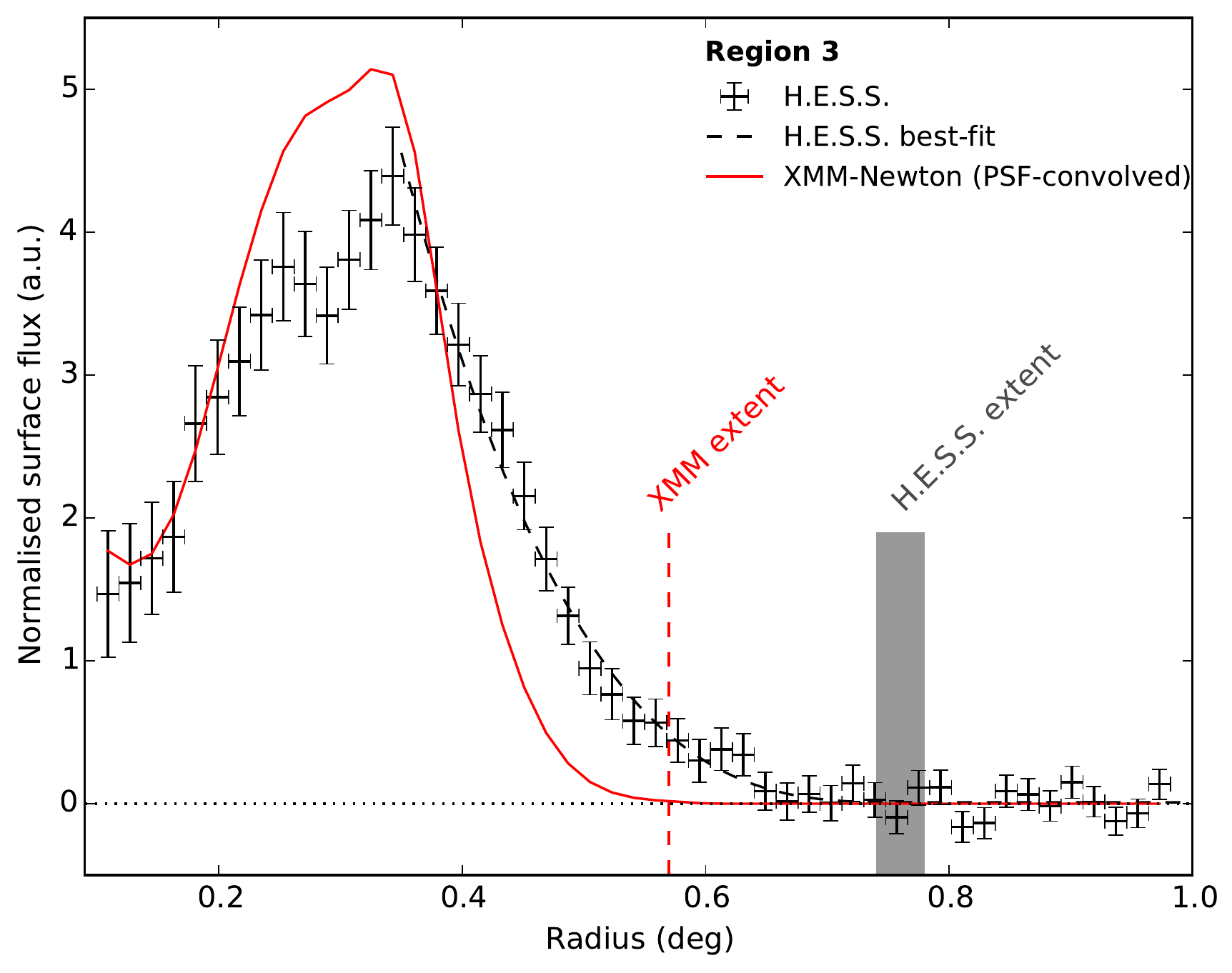}
\end{center}
\caption{\label{fig:RXJ1713_Profiles}{\bf Left}: Quadrants used in the \HESS VHE \gray\ image of RX~J1713.7-3946 to investigate possible particle escape.
{\bf Right}: Radial profile of \gray\ (black) and X-ray (red) emission in quadrant number 3~\cite{hess-rxj1713-icrc2015}.}
\end{figure}

Since the last \HESS\ publication~\cite{hess-rxj1713-deep}, the amount of data has been doubled.
At the same time,  high-resolution / high-throughput analysis techniques have been developed, e.g.~\cite{denaurois2009},
allowing to spatially resolve spectra with unprecedented resolution ($\lesssim 0.05^\circ$). The resulting \gray\ image
is shown in Fig.~\ref{fig:RXJ1713}, left, and has an energy threshold of $\sim 250\U{GeV}$. 
For the first time, maps of physical quantities, such as magnetic field in the case of a leptonic scenario,
can be produced. Fig.~\ref{fig:RXJ1713}, right, shows the distribution of magnetic field inside the SNR, inferred
from a one-zone leptonic model.
In addition, the high statistics collected and the high resolution analysis permits a detailed comparison of the morphology 
between VHE $\gamma$ rays and X-rays.
The \gray\ image from Fig.~\ref{fig:RXJ1713} has been in five quadrants (Fig.~\ref{fig:RXJ1713_Profiles}, left) for which the radial profiles have been determined.
In several regions, such as region 3 (Fig.~\ref{fig:RXJ1713_Profiles}, right) the \gray\ emission appears significantly more extended than
the X-ray one (smeared to the \HESS\ PSF), indicating for the first time diffusion of
particles outside the shell and even escape of high energy particles, most likely of hadronic origin.

Further studies, and in particular determination of the energy spectrum of escaping particles
and detailed investigation of the interaction with the surrounding medium,
would further improve our understanding of the underlying acceleration \& diffusion mechanisms.

\subsection{IC 443}

In the GeV domain, the brightest SNRs, IC~443, W44 and W51C are supernovae remnants which interact with MCs,
that provide target material for accelerated protons and nuclei.
The so-called ``{\it pion bump}'', a low energy cutoff
in the energy spectrum that clearly indicates emission of $\gamma$ rays channelled through the decay of neutral pions produced 
by hadronic interactions, was first detected in IC~443 by AGILE~\cite{2010ApJ...710L.151T}. \LAT\ recently reported its
detection in IC~443 \& W44~\cite{fermi-ic443-w44}.

\begin{figure}[t!]
\begin{center}
\includegraphics[width=0.35\textwidth]{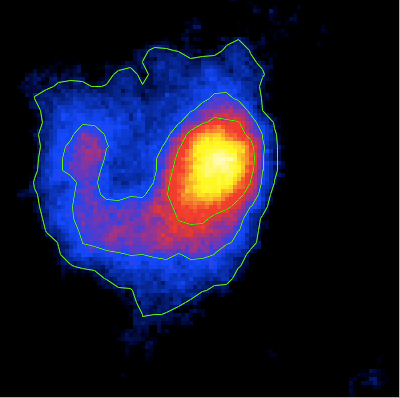}
\includegraphics[width=0.35\textwidth]{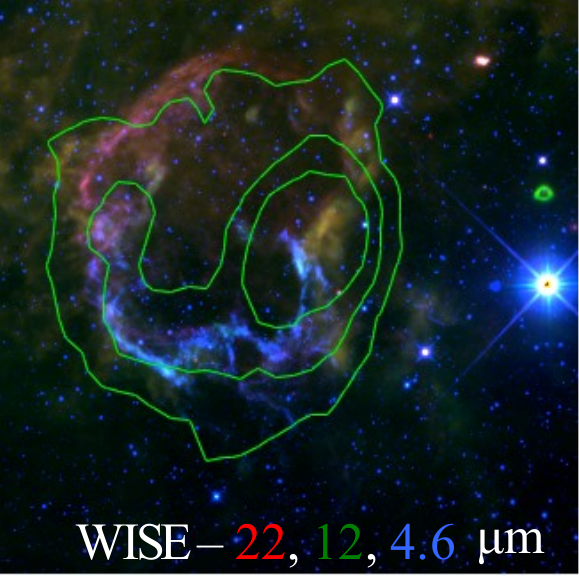}
\end{center}
\caption{\label{fig:IC443}{\bf Left:} IC 443 shell resolved by VERITAS. {\bf Right}: VERITAS contours superimposed with the 
Wide-field Infrared Survey Explorer (WISE) map in false colors~\cite{veritas-ic443-icrc2015}. WISE image credit: NASA/JPL-Caltech/UCLA}
\end{figure}

IC~443, also known as the {\it Jellyfish Nebula} is a relatively nearby ($1.5\U{kpc}$) and extended (50' across) SNR, 
and has been extensively studied for decades.
It exhibits a shell-like morphology in optical and in radio~\cite{1986A&A...164..193B}, consisting 
of two connected sub-shells with different centers and radii (Fig.~\ref{fig:IC443}, right). 
The complex morphology is attributed to the presence of sharp density
gradients and different MCs in the surroundings, e.g.~\cite{1977A&A....54..889C}, reaching very high density (up to $\sim 10\ 000 \UU{cm}{-3}$), 
The latter decelerate the blast wave and affect the expansion of the shell and its morphology.

After initial detection in VHE \grays\ by MAGIC~\cite{magic-ic443} and VERITAS~\cite{veritas-ic443},
the latter presented for the first time at this conference the shell of IC~443 resolved in $\gamma$ rays~\cite{veritas-ic443-icrc2015}.
The corresponding image, superimposed with 3, 6 and 9 $\sigma$ contours, is shown in Fig.~\ref{fig:IC443}, left, and exhibits a clear, shell-like morphology. 
This will allow for the first time to extract spectra from different regions within the remnant and to probe the environmental dependence of cosmic-ray diffusion.
The centroid of the emission is consistent with the emission previously reported by MAGIC and corresponds
to the densest part of the MC, where a OH maser emission is also found~\cite{1997ApJ...489..143C,2006ApJ...652.1288H}, and where
the blast wave is significantly decelerated 
(Fig.~\ref{fig:IC443}, right), 
thus confirming the hadronic origin of the emission.

\section{Gamma-ray binaries: Swiss clocks vs fuzzy clocks}

During the last decade, binary systems consisting of a compact object (stellar-mass black hole or neutron star) and a non-degenerate high-mass companion star
have been firmly established as VHE \gray\ emitters.
To date, five \gray\ binaries, defined as binary systems displaying most of their non-thermal emission at \gray\ energies, have been found. 
In order of detection, they are: PSR~B1259-63~\cite{hess-psrb1259}, LS~5039~\cite{hess-ls5039},
\LSI~\cite{magic-lsi61303}, HESS~J0632+057~\cite{hess-j0632} and 1FGL~J1018.6-5856~\cite{fermi-j1018}.
They exhibit vastly different periods, ranging from few days to several years, and harbour a massive O or Be star
with temperature of  $20 000 - 40 000\U{K}$. The orbital period and the companion star mass and temperature of these 
five \gray\ binaries are summarized in table~\ref{tab:Binaries}. For an extensive review, see~\cite{2013A&ARv..21...64D} and references therein.

\begin{table} [b!]
\begin{center}
\begin{tabular}{l|c|c|c}
\hline
& Period (days)	& $M_\star (M_\odot)$ & $T_\star (K)$ \\
\hline
PSR B1259-63	&1236	&31 & 33500\\
LS 5039		&3.9	&23 & 39000\\
\LSI	&26.4	&12 & 22500\\
HESS J0632+057	&315	&16 & 30000\\
1FGL J1018.6-5856	&16.6	&31 & 38900\\
\hline
\end{tabular}
\caption{\label{tab:Binaries}Orbital period and companion star mass and temperature of the five \gray binary systems known so far. From ~\cite{2013A&ARv..21...64D} and references therein.}
\end{center}
\end{table}

\begin{figure}[b!]
\begin{center}
\includegraphics[width=0.45\textwidth]{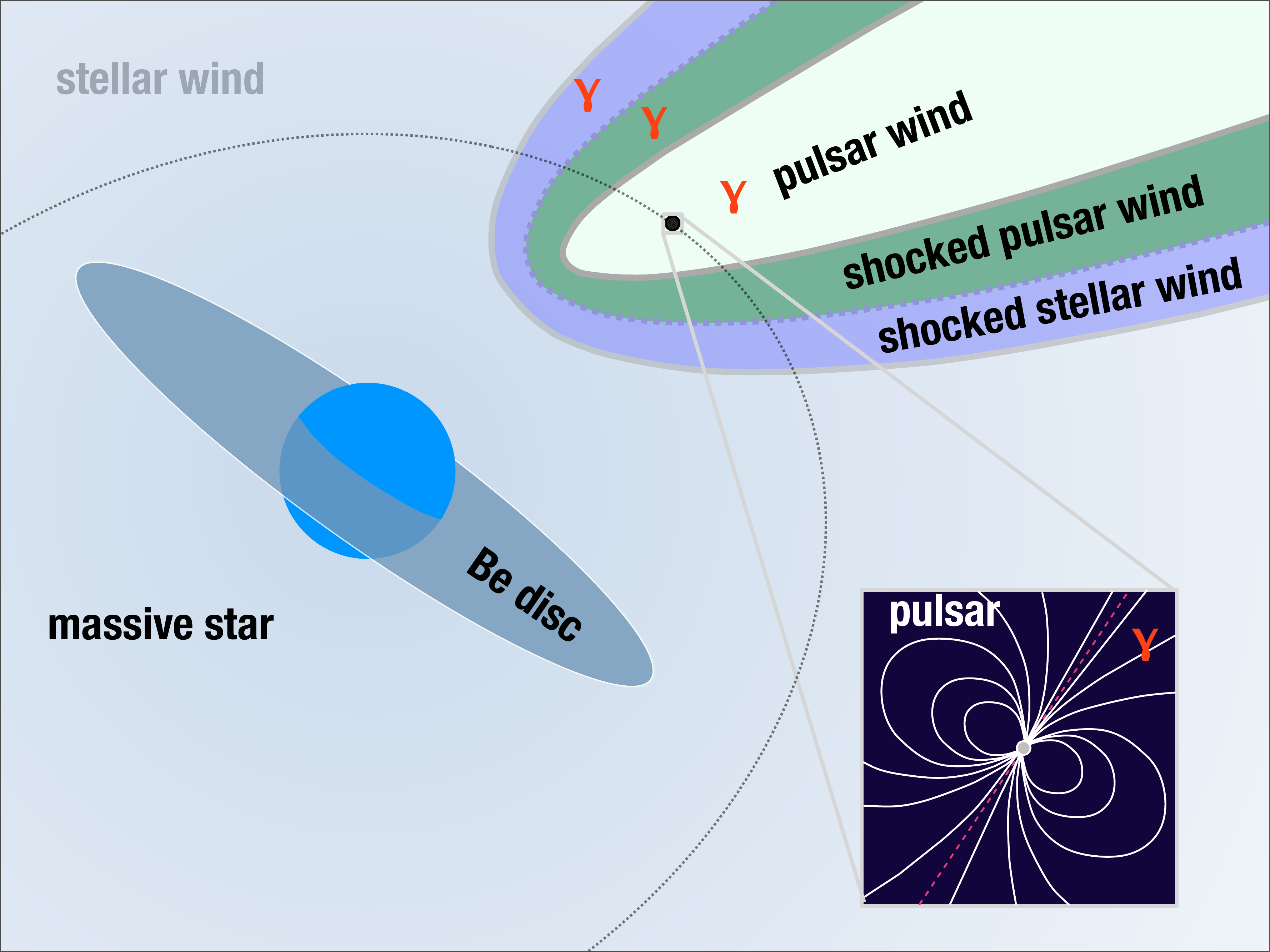}
\includegraphics[width=0.45\textwidth]{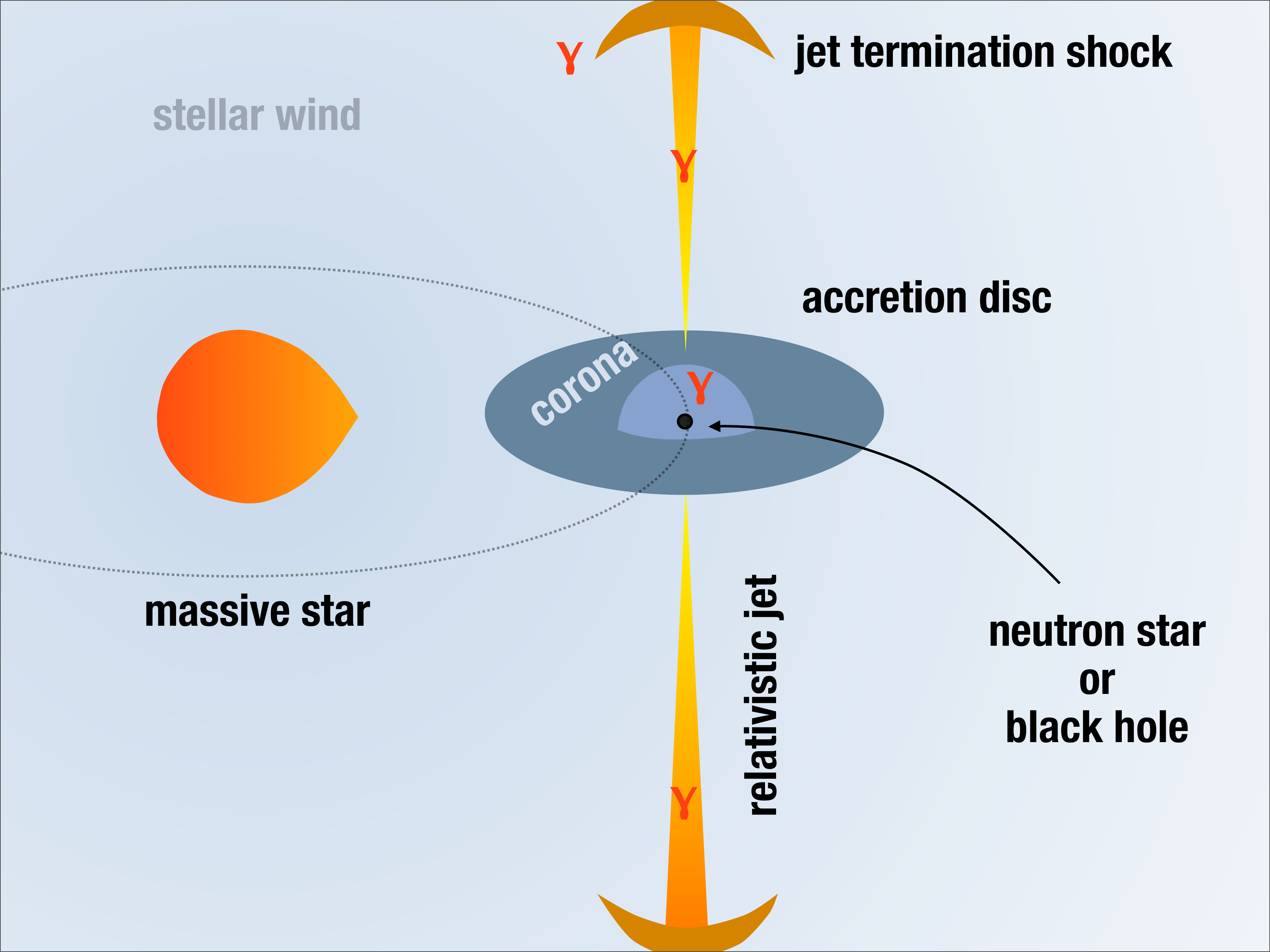}
\end{center}
\caption{\label{fig:BinariesGeometry}The two main scenarios for \gray\ emission from binaries.
{\bf Left}: in the pulsar wind scenario, the variable emission arises from the interaction of
the  pulsar wind with the strong stellar wind of the companion star.
{\bf Right}: in the microquasar scenario, the emission is powered by the accretion of the massive star onto
the compact object (black hole or neutron star) giving rise to relativistic jets.
Reproduced from~\cite{2013A&ARv..21...64D}.}
\end{figure}

Two main scenarios are currently considered to explain the variable emission from \gray\ binaries (Fig.~\ref{fig:BinariesGeometry}).
In the pulsar-wind scenario (left), the system is powered by the wind of a high spin-down power pulsar, that interacts with the wind
of the massive star. In the microquasar scenario (right), the system is powered by the accretion of the massive star onto the compact object.
Below, we briefly summarize some of the main properties which make \gray\ 
binaries unique laboratories for high-energy astrophysics:

\begin{itemize}
\item They operate in rather extreme environment, with very dense radiation fields and stellar winds, 
that are not easily found in other systems.
These radiation fields can induce strong absorption of $\gamma$ rays by pair creation on the stellar photons, 
resulting in a regular modulation given by the periodic changes in the geometry of the system~\cite{hess-ls5039-paper2,dubuscerutti2008}.
\item The physics at play is complex, both in accretion-based or wind-wind interaction scenarios, with $\gamma$ rays thought to 
originate in the interactions with strongly anisotropic radiation fields and/or surrounding material.
The conditions, especially in highly eccentric systems, vary  significantly along the orbit, resulting in competing
effects: acceleration efficiency, energy losses, radiative output and absorption can vary by orders of magnitude, thus producing
complex and different phenomenologies. A detailed knowledge of the system geometry is a prerequisite for any attempt to understand the
emission mechanisms. On the other hand, this offers the unique possibility to test different physical conditions in a single system.
In this sense, \gray\ binaries are good laboratories for testing acceleration and radiation/absorption mechanisms. 
\item The binary nature of the systems allows some of their properties (masses, or at least mass function, 
inclination, orbit) to be measured with a good precision, allowing rather accurate modelling.
\end{itemize}

\subsection{PSR~B1259-63 and LS~5039 - The Swiss clocks}

PSR~B1259-63 is a long period ($3.4\U{years}$) binary system in a highly eccentric orbit~\cite[and references therein]{hess-psrb1259}.
It is the only \gray\ binary where the nature of the compact object -- a pulsar with a spin period of $48\U{ms}$ -- has been firmly established
through observations far away from  periastron, where the radio pulses are not absorbed by the stellar wind.

Significant \gray\ emission arises only close to periastron, when the pulsar wind interacts with the stellar and circumstellar wind, giving
rise to an enhancement of the non-thermal emission visible across the entire electromagnetic spectrum. In addition, the pulsar crosses the circumstellar
disk twice, $\sim 20$ days before and after periastron~\cite{Chernyakova2014}.

\begin{figure}[ht!]
\begin{center}
\includegraphics[width=0.7\textwidth]{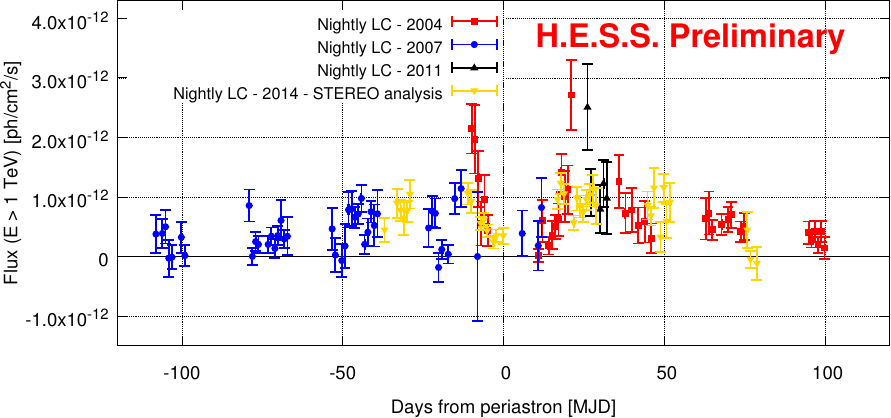}\\
\vspace{1em}
\includegraphics[width=0.7\textwidth]{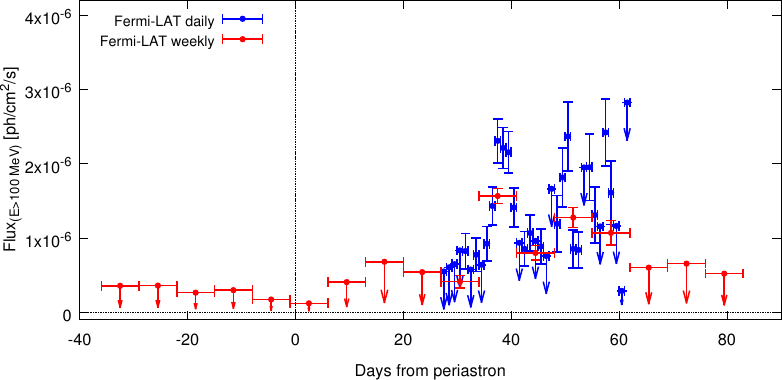}
\end{center}
\caption{\label{fig:HESS_PSRB1259} {\bf Top:} \HESS\ daily light curves of PSR B1259-63, folded over several periastron passages.
{\bf Bottom:} Flux obtained from the analysis of \LAT\ data in the energy range $100\U{MeV} - 300\U{GeV}$. Note that the horizontal scales
are different~\cite{hess-psrb1259-icrc2015}.}
\end{figure}

VHE emission from PSR~B1259-63 was first detected by \HESS\ during the periastron passage in 2004~\cite{hess-psrb1259}, and
was extensively monitored during the 2007 and 2010/2011 passages~\cite{hess-psrb1259-paper2,hess-psrb1259-paper3}, whereas the source was
not detected during observations far from periastron.

During the 2010/2011 passage, \LAT\ detected a high energy (HE) flare $\sim 30$ days after the periastron passage~\cite{fermi-psrb129}. This flare,
without any counterpart at other wavelengths, came as a complete surprise to the community and its nature remains controversial.

Observations with \HESS\ during the 2014 periastron passage were reported at the conference~\cite{hess-psrb1259-icrc2015}.
The overall light-curve (Fig.~\ref{fig:HESS_PSRB1259}, upper panel) is generally consistent with the previous (2004, 2007 and 2010/2011) periastron passages,
with, however, some evidence for variability on short time-scales. As in 2010/2011, a strong enhancement of the HE \gray\ flux was measured by \LAT\ around
30 days after periastron (Fig.~\ref{fig:HESS_PSRB1259}, lower panel), with a double-peak structure, indicating that this puzzling HE flare is probably
a recurrent feature of the system.
For the first time, the large \HESS-II telescope, CT5, was operating during the 2014 periastron passage. 
No counterpart of the \LAT\ flare was found in the monoscopic data, down to $200\U{GeV}$.
However, the source appeared rather bright even $\sim 50$ days after periastron, displaying a relatively 
high flux state contemporaneous to the development of the HE flare.

LS~5039 is a compact binary system, with a period of $3.9$ days. It has been extensively monitored 
during the last 10 years by \HESS~\cite{hess-ls5039,hess-ls5039-paper2}
and \LAT~\cite{fermi-ls5039}. A strong modulation of the HE and VHE fluxes with the orbital period, 
with anti-correlation between the two energy bands, is attributed mainly
to anisotropic IC emission and variable absorption of the VHE \gray\ flux by pair 
creation on stellar photons (the absorption being maximal  at the superior conjunction, when the 
compact object is behind the star) and subsequent production of HE photons by cascading~\cite{dubuscerutti2008}.

More recent data~\cite{hess-ls5039-icrc2015}, including \HESS-II observations, confirms the perfectly regular behaviour of the source over more than 10 years,
and contribute to close the spectral gap between \LAT\ and \HESS.

\subsection{\LSI\ - The fuzzy clock}

The binary system \LSI\ is composed of a Be star~\cite{1981PASP...93..486H} with a circumstellar disk and an 
unidentified compact object (neutron star or black hole), in a highly eccentric orbit
($e=0.54 \pm 0.03$). It was first detected by MAGIC~\cite{magic-lsi61303} and later on confirmed by VERITAS~\cite{veritas-lsi61} and 
\LAT~\cite{fermi-lsi61303}. Although the VHE emission appears to be linked
with the orbit (with significant periodicity at VHEs), unlike LS~5039, the orbital phases of VHE detection have varied considerably in different observation campaigns.
In early observations, the emission was confined to the phase $0.5 \leq \phi \leq 0.9$~\cite{veritas-lsi61,veritas-lsi61-mwl,magic-lsi61303-paper3}. 
In observations performed in 2010 the source was instead detected at orbital phases $0 \leq \phi \leq 0.1$~\cite{veritas-lsi61-2011}.

In addition, the non-thermal emission is characterized  by a super-orbital modulation ($1667\pm 8$ days), detected in radio~\cite{2002ApJ...575..427G}
and HE $\gamma$ rays~\cite{fermi-lsi61303-superorbital}. Overall, \LSI\ is an unusual accelerator with rapidly changing conditions. Among the possible
explanations for the variability,  turbulent mixing of stellar and pulsar winds, a structured stellar wind~\cite{2015A&A...574A..77P}, or the
interaction of pulsar winds with stellar disk have been considered.

The exact nature of the system and the underlying emission mechanisms are still subject of debate, with possible scenarios including:

\begin{itemize}
\item a pulsar wind scenario, similar to that proposed for LS~5039 and PSR~B1259-63, and supported 
by the detection by VLA of rotating cometary-like structures~\cite{Dhawan2006},
\item a microquasar scenario, supported by the detection with MERLIN of extended jet-like structures~\cite{massi2004},
\item finally, after a very short burst ($<0.1\U{s}$) detected with Swift, \LSI\ has been proposed to 
be the first binary system comprising a magnetar~\cite{2012ApJ...744..106T}.
\end{itemize}

\begin{figure}[b!]
\begin{center}
\includegraphics[width=0.49\textwidth]{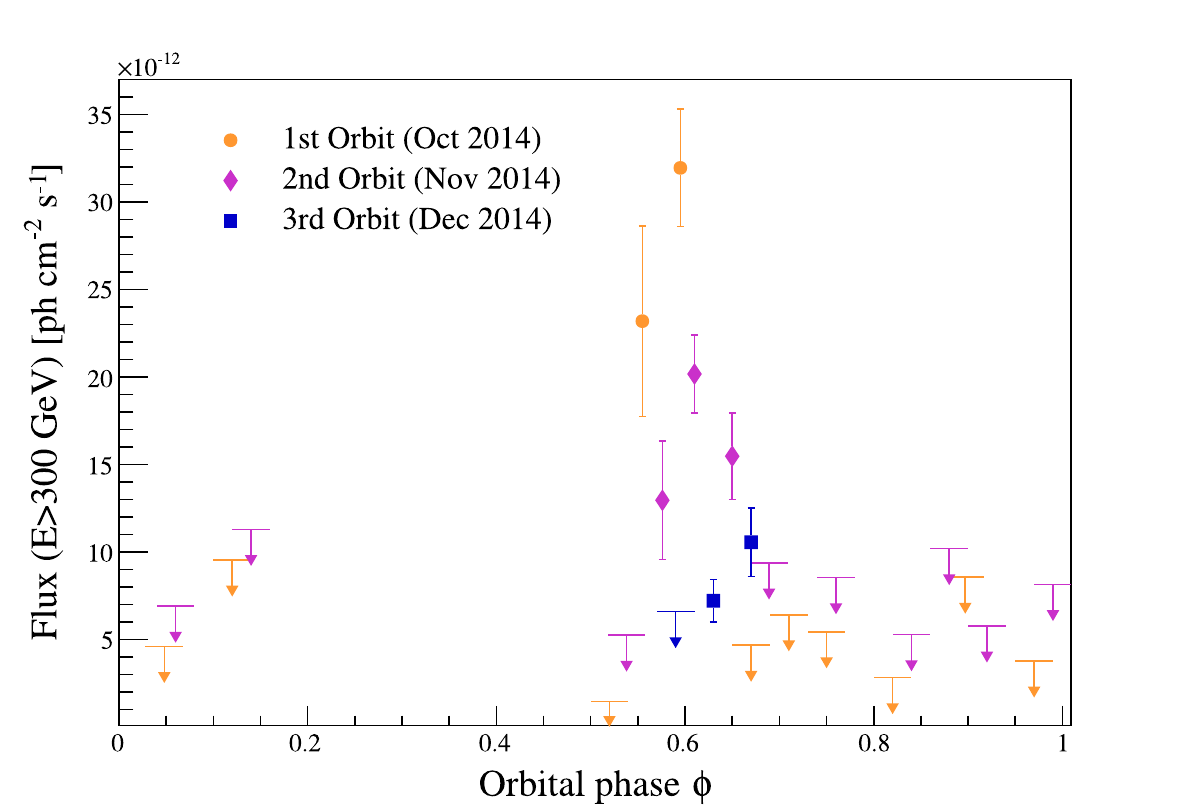}
\includegraphics[width=0.49\textwidth]{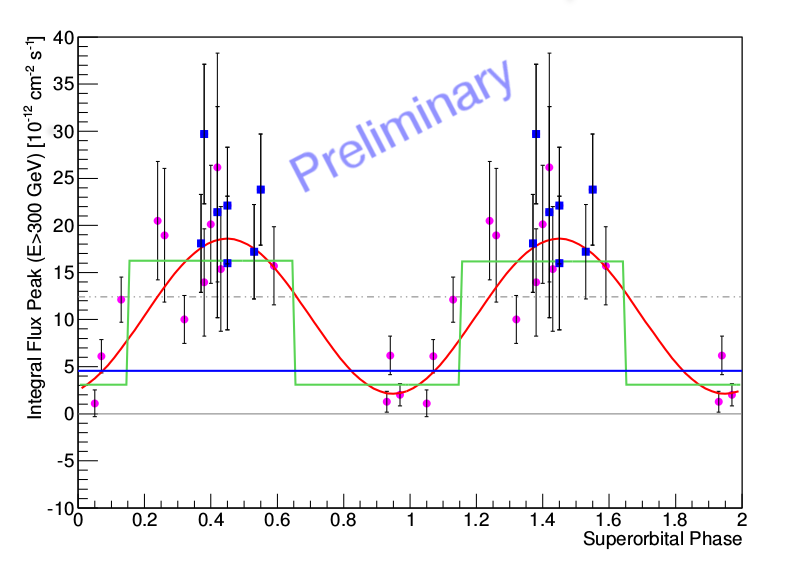}
\end{center}
\caption{\label{fig:LSI}{\bf Left:} VERITAS nightly light curve of \LSI\ during the 2014 observation season~\cite{veritas-lsI-icrc2015}.
{\bf Right:}MAGIC peak flux of \LSI\ in phase $0.5-0.75$, folded with the super-orbital period~\cite{magic-binaries-icrc2015}.}
\end{figure}

Long-term observations are key to understand the variability of particle acceleration and \gray emission in \LSI. After 8 years of monitoring,
VERITAS acquired 160 hours of deep observations of the system. 
In October 2014, an exceptionally bright ($> 25\%$ of the Crab Nebula flux) and fast ($<1$-day rise time) flare was detected by VERITAS~\cite{veritas-lsI-icrc2015}.
The corresponding light-curve is shown in Fig.~\ref{fig:LSI}, left. The system retreated to its quiescent level shortly after, in December 2014.
In contrast to the flaring episode of 2007, no significant enhancement  was detected at the same time in \LAT, suggesting that different populations of particles
might be at the origin of the HE and VHE emissions.

Several mechanisms have been proposed to explain the irregular behaviour of \LSI. In particular, the 
so-called {\it flip-flop} mechanism~\cite{2012ApJ...744..106T,2012ApJ...744L..13L} has been put forward,
which relies on a periodic change of the mass-loss rate of the star:
close to the periastron, the pulsar magnetosphere is disrupted by the circumstellar wind and the system enters the ``{\it propeller}'' regime,
characterized by a suppression of the VHE \gray\ emission. At other phases, and in particular around apastron, the usual ``{\it ejector'}'' regime
applies, with acceleration of particles up to TeV energies in the pulsar magnetosphere. In that scenario, the size of the circumstellar disk would
change according to the massive star mass-loss rate. For high mass-loss rate, the ``{\it propeller}'' regime could then span the full orbit.

In order to investigate the super-orbital variability of \LSI , MAGIC performed extended observations between 2010 and 2014~\cite{magic-binaries-icrc2015}. 
These, together with the archival data, resulted in the folded light-curve of Fig.~\ref{fig:LSI}, right, which is consistent with a sinusoidal variation at the super-orbital
phase, and excludes a constant flux.
Contemporaneous observations of the $\mathrm{H}_\alpha$ line emission with the LIVERPOOL telescope near the MAGIC site could not, however, establish any
correlation with the mass-loss rate and the super-orbital phase, leaving the debate still open.


\section{Very High Energy pulsars}

\begin{figure}[h!]
 \begin{minipage}[b]{0.49\textwidth}
   \centering
   \includegraphics[width=\textwidth]{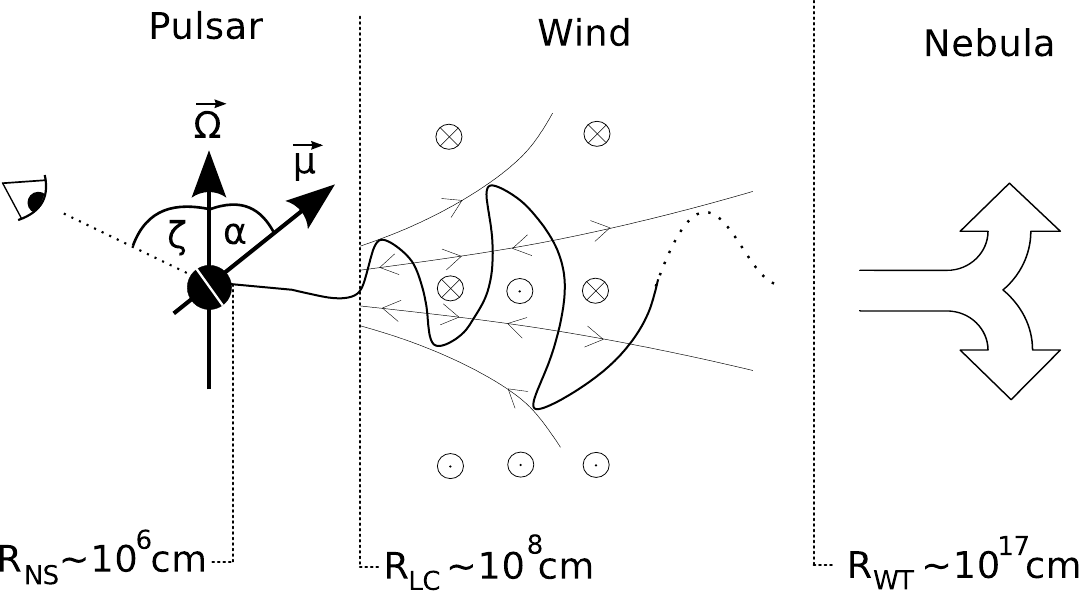}
  \end{minipage}
\hfill
  \begin{minipage}[b]{0.47\textwidth}
   \centering
   \includegraphics[width=\textwidth]{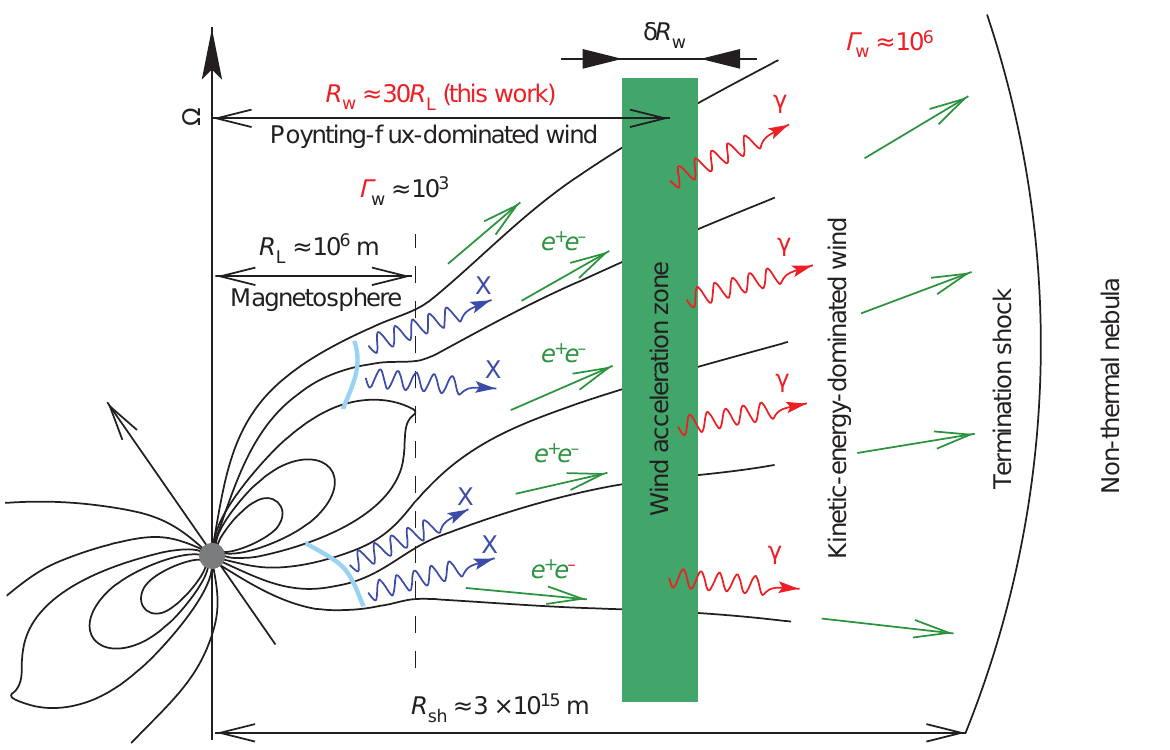}
  \end{minipage}
\caption{\label{fig:Pulsars}
{\bf Left:} Sketch of the components of a PWN.
The equatorial current sheet (thick straight line) and the magnetic fields (thin dashed
line and circles) are shown in the wind region. Reproduced from~\cite{2014RPPh...77f6901B}.
{\bf Right:} Sketch of the different regions surrounding the pulsar, including the magnetosphere, 
the unshocked wind region and the PWN.
In the wind acceleration region (in green), the electromagnetic energy
contained in the pulsar wind is converted into bulk kinetic energy of a relativistic e+/e- plasma.
Particles are isotropized at the termination shock, and release their energy
through synchrotron and IC radiation, resulting in the formation of a non-thermal emitting nebula.
Reproduced from~\cite{2012Natur.482..507A}.
}
\end{figure}

The first detection of VHE \gray\ pulsed emission above $25\U{GeV}$ of the Crab pulsar by the MAGIC collaboration~\cite{magic-crab},
and the later detection above $100\U{GeV}$~\cite{veritas-crab,magic-crab-2012}, triggered an intense theoretical and experimental activity.
Energetic pulsars are complex systems, comprising the neutron star itself, its magnetosphere, the ultra-relativistic wind and the
PWN surrounding the pulsar (Fig.~\ref{fig:Pulsars}, left).

\subsection{Crab Pulsar}

The Crab Pulsar, PSR~B0531+21, is a young energetic pulsar, with a rotation period of $\sim 33 \U{ms}$ and 
a spin-down power of $4.6\times 10^{38} \U{erg}\UU{s}{-1}$,
and is the remnant of the historical supernova in 1054 A.D. Being relatively nearby ($\sim 2\U{kpc}$), 
it is one of the most studied non-thermal astronomical sources.
Pulsed HE \gray\ emission was already detected by EGRET~\cite{1996A&AS..120C..61N}, triggering 
an intense and successful search for pulsed emission in VHE $\gamma$ rays:

\begin{itemize}
\item 2008: First detection of pulsed emission above $25\U{GeV}$  by the MAGIC collaboration~\cite{magic-crab}
\item 2011: First detection of pulsed emission in the $120-400\U{GeV}$ energy range by the VERITAS collaboration. 
Pulsed emission in this energy range was theoretically rather unexpected and started to challenge existing pulsar models.
The VHE emission appeared to originate from a new, power-law component, possibly extending 
to even higher energies. In order to avoid strong energy losses of high energy electrons by synchrotron radiation, these $\gamma$ rays must 
be produced more than 10 stellar radii away from the neutron star~\cite{veritas-crab}. 
\item 2011: First spectrum in the $25-100\U{GeV}$ energy range by MAGIC~\cite{magic-crab-2011}.
\item 2012: Spectrum extending to $400\U{GeV}$ by MAGIC~\cite{magic-crab-2012}
\item 2014: Detection of VHE emission from the bridge between the two peaks P1 and P2 at energies $\geq 50\U{GeV}$ by MAGIC~\cite{magic-crab-2014}
\end{itemize}

\begin{figure}[b!]
 \begin{minipage}[b]{0.53\textwidth}
   \centering
   \includegraphics[width=\textwidth]{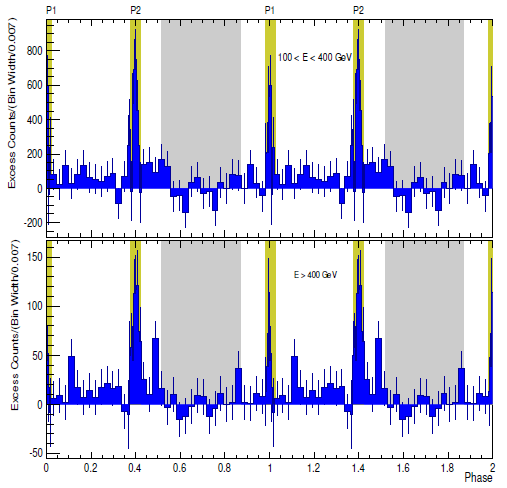}
  \end{minipage}
  \begin{minipage}[b]{0.46\textwidth}
   \centering
   \includegraphics[width=\textwidth]{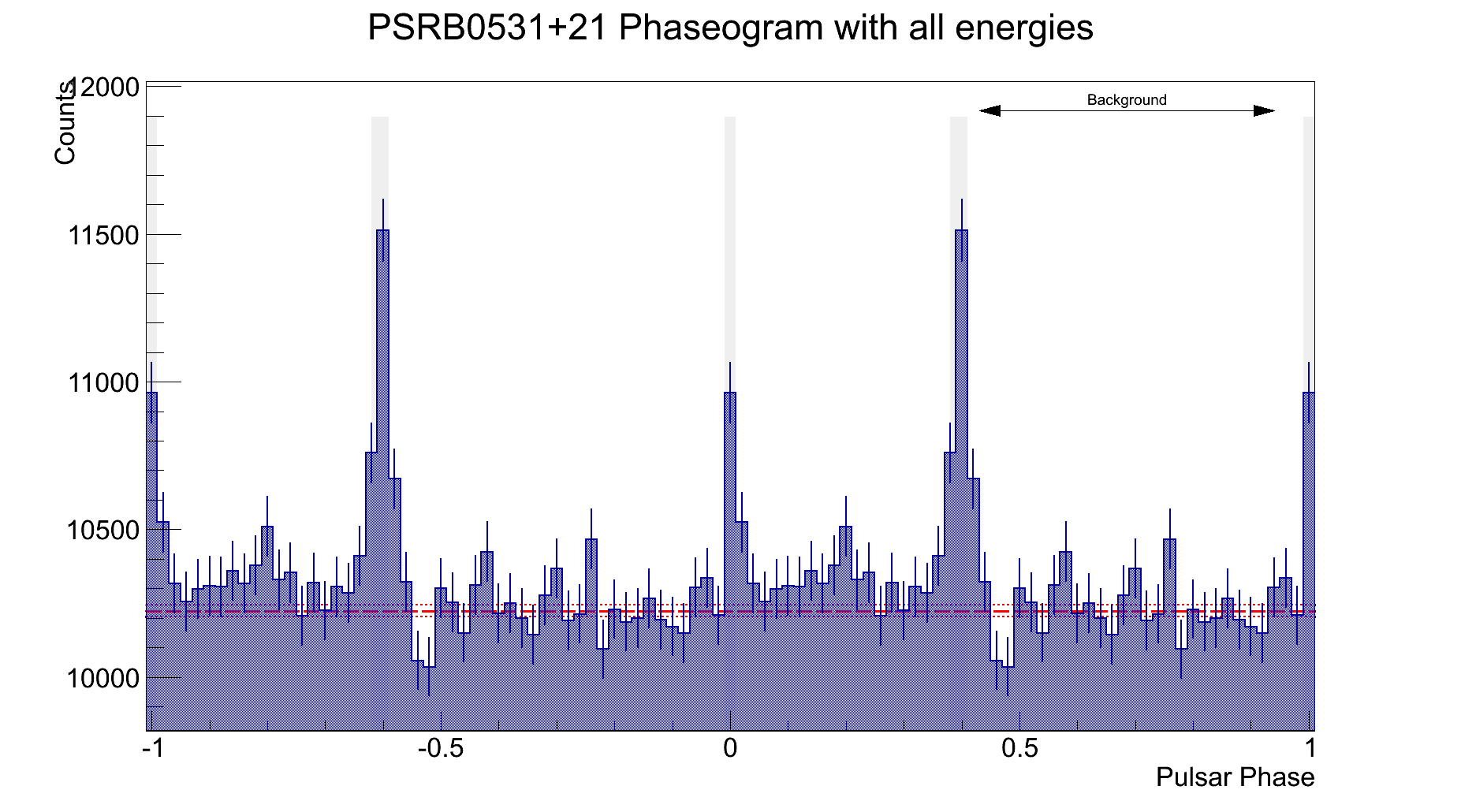}\\
   \includegraphics[width=\textwidth]{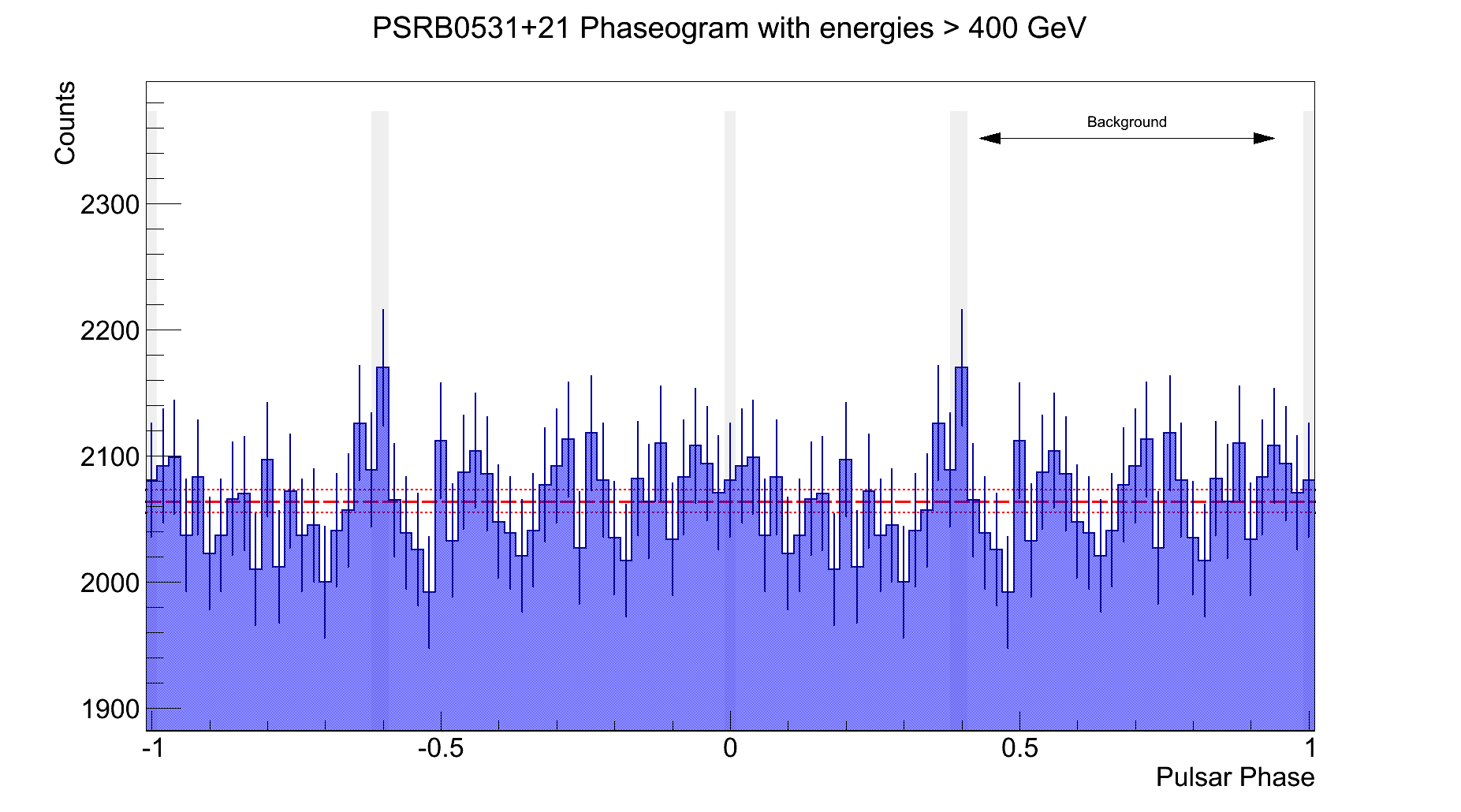}
  \end{minipage}

\caption{\label{fig:CrabPhasogram}Phase-folded distribution of events from the Crab Pulsar(``{\it phasogram}'').
{\bf Left:} MAGIC results in the $100-400\U{GeV}$ energy domain (top) and $>400\U{GeV}$ (bottom)~\cite{magic-crab-icrc2015}.
{\bf Right:} VERITAS results for all energies (top) and $>400\U{GeV}$ (bottom)~\cite{veritas-crabpulsar-icrc2015}.}
\end{figure}

The detection of VHE pulsed emission of the Crab Pulsar triggered a lot of theoretical activity. Several models have been proposed,
all of them placing the emission far away from the neutron star, close to the light cylinder or even beyond.
In the Annular Gap model for instance~\cite{2012ApJ...748...84D}, the gap region is located between the critical 
field lines and the last open field lines,
possibly extending up to the light cylinder. 
As for the outer gap~\cite{2012ApJ...744...34V} or slot gap models~\cite{2008ApJ...680.1378H}, VHE emission arises from IC up-scattering 
of lower energy photons, see e.g.~\cite{magic-crab-2011}. VHE \gray\ emission could also be produced in the striped wind of the pulsar,
from IC scattering of pulsed X-ray photons~\cite{2000MNRAS.313..504B,2012Natur.482..507A,2012MNRAS.424.2023P} (Fig.~\ref{fig:Pulsars}, right). 
For a recent review of existing models, see for instance~\cite{2014RPPh...77f6901B}.


Both the VERITAS and MAGIC collaborations have devoted a large amount of observation time (respectively $195$ and $320$ hours) 
to further investigate the energy extension of the emission and evolution of pulse shape with energy, 
for better constraining the existing emission models.
The phasograms of the Crab Pulsar obtained by the MAGIC and VERITAS collaborations are shown in Fig.~\ref{fig:CrabPhasogram}~\cite{magic-crab-icrc2015,veritas-crabpulsar-icrc2015}.
Both of them reported the detection of P1 and P2 peaks in the $100-400\U{GeV}$ energy domain at a high significance level. 
Above $400\U{GeV}$, only MAGIC detects the two peaks, whereas the combined excess counts
from P1 and P2 yields only $2\,\sigma$ for VERITAS. MAGIC reported at this conference pulsed emission up to approximately $2\U{TeV}$, 
which requires a parent population of electrons with a Lorentz
factor of at least $5\times 10^6$. Above $950\U{GeV}$, $119\pm 54$ and $190\pm 56$ excess events are detected in pulses P1 and P2 respectively, 
corresponding to significances of $2.2$ and $3.5\,\sigma$~\cite{magic-crab-2015}. This exciting discovery
further challenges pulsar models, with VHE \gray\ emission likely to take place in the neighbourhood  of the light cylinder ($r \sim 1600\U{km}$).
VERITAS could not yet confirm this important discovery. Nevertheless, further observation extending up to $300\U{hours}$ have been 
planned by the VERITAS collaboration to extend results to higher energies.

MAGIC also recently reported the detection of emission from the bridge between P1 and P2 above $50\U{GeV}$, 
with a statistical significance of $6.2\,\sigma$~\cite{magic-crab-2014}.
Although qualitative explanations of the bridge can be found, for instance in~\cite{2012MNRAS.424.2079B},
a consistent picture of the  VHE Crab pulsed emission is still lacking further theoretical developments.


\subsection{Vela Pulsar}

Despite its longer period, $P = 89.3\U{ms}$, compared to the Crab pulsar, its larger characteristic age $\tau_c = 11\U{kyr}$ and its much smaller
spin-down power $\dot E  = 6.9 \times 10^{36} \U{erg} \UU{s}{-1}$~\cite{2005AJ....129.1993M}, the Vela pulsar is,
due to its proximity ($d = 287^{+19}_{-17}\U{pc}$~\cite{2003ApJ...596.1137D})
exceptionally bright in radio and in HE $\gamma$ rays, and is, after the Crab pulsar, one of the best natural candidates for VHE emission.
The Vela pulsar is surrounded by a strong wind nebulae, detected in particular by \LAT~\cite{2010ApJ...713..146A}.

\begin{figure}[b!]
\begin{center}
\includegraphics[width=0.8\textwidth]{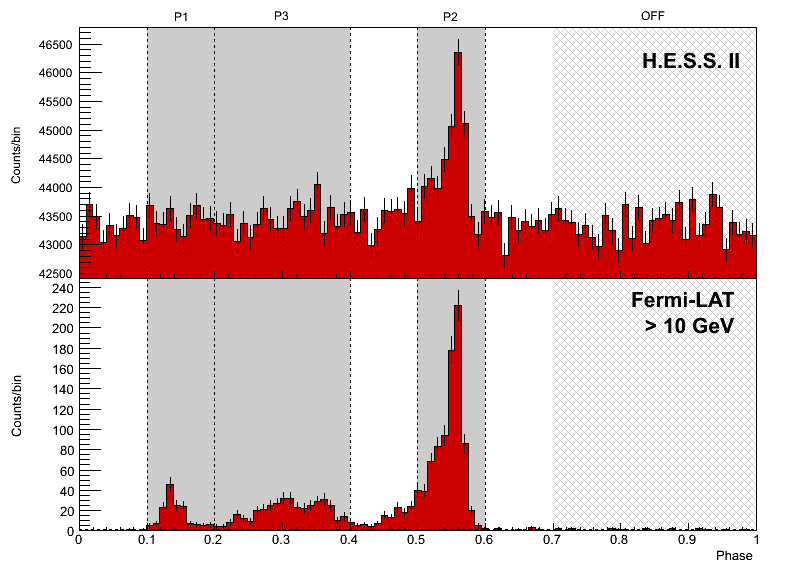}
\end{center}
\caption{\label{fig:VelaHESS}Phase-folded distribution of events of the Vela Pulsar with \HESS-II (top) and \LAT (bottom)~\cite{hess-vela-icrc2015}.}
\end{figure}

The HE \gray\ phasogram is characterized by two main sharp peaks (P1 and
P2) and a third peak (P3) in the bridge~(Fig.~\ref{fig:VelaHESS}, bottom). 
The ratio of the peak intensity between P1 and
P2, and the location and intensity of P3  vary with energy~\cite{2010ApJ...713..154A,2010ApJS..187..460A}.


Various models are able to properly describe the HE \gray\ emission of the Vela pulsar.
For instance, in the framework of the annular model, e.g.~\cite{2011ApJ...731....2D},
curvature radiation of primary particles is responsible for the emission, without the need 
of curvature radiation from secondary particles nor IC scattering.

The pulsed flux component does however fall very quickly with energy, and the lack of statistics with the \LAT\ 
does not allow the shape of the cut-off to be constrained, nor the  presence
of another component at higher energies, possibly similar to that observed in the Crab pulsar, to be identified. 
\HESS-II, with is much larger effective area ($\sim 10^4 \UU{m}{2}$ at $20\U{GeV}$) and low energy threshold, is perfectly suited to the 
observation of VHE pulsars. $24\U{hours}$ of high quality data were obtained by \HESS-II in 2013 and 2014,
and were analyzed with a dedicated set of analysis cuts, leading to a very low energy threshold ($\sim 20\U{GeV}$).

The phased-folded event distribution obtained is shown in Fig.~\ref{fig:VelaHESS}, top. Nearly $10\ 000$ excess events
were detected in the pulse P2, corresponding to a Li \& Ma significance of $12.8\,\sigma$. No significant excess is observed from
the P1 peak, nor from the P3, which is in agreement with an extrapolation of the \LAT\ phasogram.

\begin{figure}[ht!]
\begin{center}
\includegraphics[height=5.5cm]{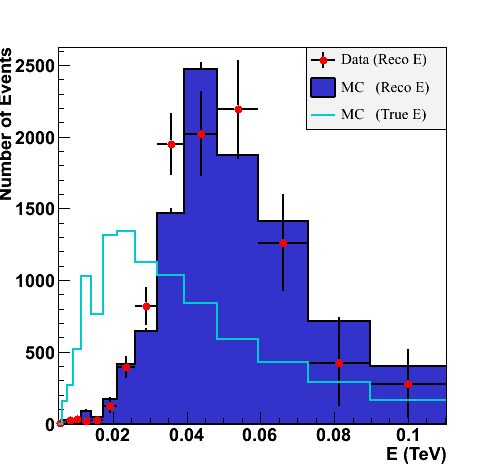}
\includegraphics[height=5.5cm]{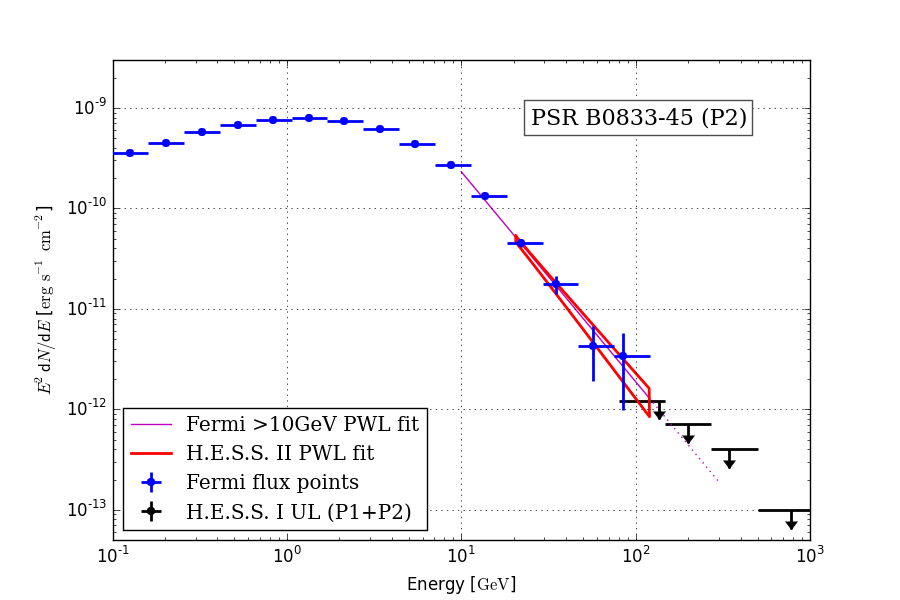}
\end{center}
\caption{\label{fig:VelaEnergyHESS}{\bf Left:} Reconstructed energy distribution of the excess events in P2 peak (red points) compared to the prediction of a Monte Carlo simulation
(dark blue histogram). The corresponding true energy distribution for the MC (light blue) indicates that \HESS-II is able to detect $\gamma$ rays with energy as low as $10\U{GeV}$.
{\bf Right:} Spectral energy distribution of the P2 peak of Vela with \LAT\ and \HESS-II~\cite{hess-vela-icrc2015}.}
\end{figure}

The energy distribution (Fig.~\ref{fig:VelaEnergyHESS}, left) of the reconstructed events from P2 is fully compatible with expectations
from Monte Carlo (MC) simulations (power law with spectrum index -4.1), and indicates that \HESS-II is able to detect $\gamma$ rays with 
energy as low as $10\U{GeV}$, although with a significant reconstruction energy bias. A preliminary spectral energy distribution
between $20$ and $120\U{GeV}$ (Fig.~\ref{fig:VelaEnergyHESS}, right)  is consistent with extrapolations from the \LAT\ measurement, 
but the available statistics and systematic uncertainties, still under study, limit the possibility to discriminate between different 
shapes of the high energy cut-off, as well as to establish a possible new component at very high energies.

\section{Powerful Stellar-like emitters in the LMC}

The Large Magellanic Cloud (LMC) is a satellite galaxy of the Milky Way, situated at $50\U{kpc}$ from 
the Earth with a size of  $\sim 10^\circ$
in the sky (corresponding to an intrinsic size of $8\U{kpc})$.
Host of the last nearby supernova (SN~1987A~\cite{1989ARA&A..27..629A}), it has been subject of extensive monitoring campaigns during the last years, in particular to investigate
how fast young SNRs can start accelerating particles to very high energies.
The LMC is also a somewhat extreme laboratory for the study of particle accelerators, due to very large cosmic ray and radiation densities, as well as intense
star formation activity and supernova rate (larger by a factor of $\sim 8$ than in the Milky Way~\cite{2009AJ....138.1243H}). It also contains the best
example of a local starburst, the Tarantula Nebula (30 Doradus), the largest
star-forming region in the Local Group of galaxies.

\begin{figure}[b!]
\begin{center}
\includegraphics[width=0.8\textwidth]{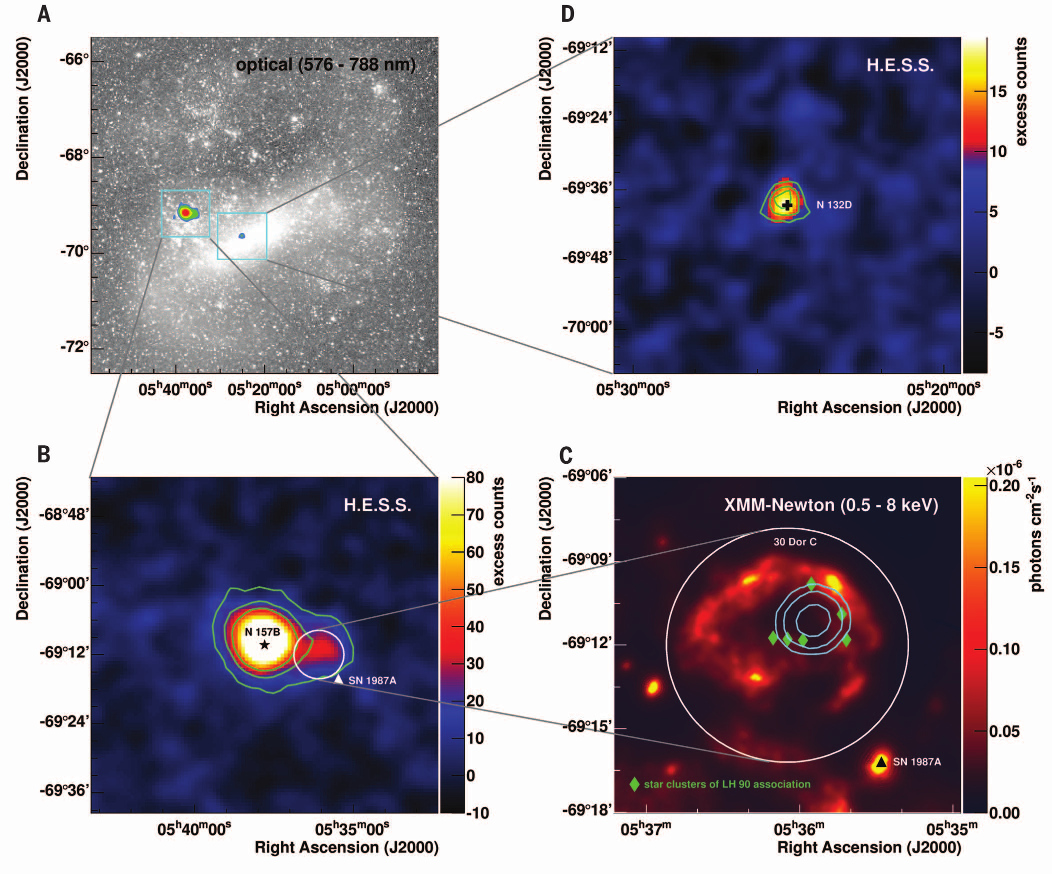}
\end{center}
\caption{\label{fig:HESS_LMC}Powerful emitters in the LMC as seen by \HESS\ and XMM-Newton~\cite{hess-lmc-science,hess-lmc-icrc2015}}
\end{figure}

In 2010 \LAT\ published~\cite{fermi-lmc} the detection of large-scale diffuse emission, attributed to the interaction 
of cosmic rays with the interstellar medium,
but could not resolve any single source in the LMC. The HE \gray\ emission appears to exhibit a much closer correlation with star forming 
regions than with interstellar gas, suggesting a 
relatively short GeV cosmic-ray proton diffusion length compared to that in the Milky Way. 

The \HESS\ collaboration accumulated, over the last years, a deep exposure of 210 hours on the LMC, targeted at the Tarantula nebula.
Sophisticated analysis techniques were used to improve the angular resolution of the instrument and resolve the different sources.
Three sources of different types were detected, all of them having some peculiar characteristics (Fig.~\ref{fig:HESS_LMC}):

\begin{itemize}
\item N157~B, already detected with a smaller data set~\cite{hess-n157B}, is a PWN associated with the most powerful pulsar known
so far (PSR~J0537-6910, spin-down power $\dot E  = 4.9 \times 10^{38} \U{erg}\UU{s}{-1}$~\cite{2014RPPh...77f6901B}).
Despite a rather low acceleration efficiency and low inferred magnetic field ($\sim 45 \U{\mathrm{\mu}G}$), it is ten times brighter in TeV $\gamma$ rays
than the Crab Nebula, most likely
due to the very high surrounding target radiation fields.
\item N~132D is a particularly old, \gray\ emitting, radio-loud SNR, possibly interacting with dense, 
shocked interstellar clouds. A hadronic
emission scenario would imply a very high energy in protons ($\sim 10^{52}\U{erg}$), 10 times larger 
than for archetypal young SNRs in the Galaxy such as Cas A.
With an age of $\sim 6000 \U{yrs}$~\cite{1998ApJ...505..732H},  between GeV and TeV emitting SNRs,
it allows probing the time span during which the SNRs are able to accelerate 
particles to tens of TeV energies.
\item The superbubble 30 Dor C is the first unambiguous detection of a superbubble in VHE $\gamma$ rays. 30~Dor~C is, with a radius of 47 pc,  
the largest non thermal, X-ray shell known so far~\cite{2004ApJ...602..257B},
filled with hot gas and likely blown into the interstellar medium by multiple supernovae and stellar winds. 
The TeV luminosity of 30~Dor~C implies extreme conditions, both for a leptonic
or a hadronic scenario, and supports the hypothesis that superbubbles might be able to accelerate particles above $10^{15}\U{eV}$.
\end{itemize}

Last but not least, the unique object SN~1987A is, despite theoretical prediction~(e.g. ~\cite{2011ApJ...732...58B}), 
not detected, which constrains the theoretical framework of particle acceleration in very young SNRs.
For first time individual, stellar-like, cosmic-ray accelerators 
are identified in an external galaxy. For the first time
as well, superbubbles are firmly identified as sources of VHE $\gamma$ rays.

\section{Peculiar extragalactic objects}

Active Galactic Nuclei, consisting of an accreting supermassive black hole associated 
with large scale, non-thermal jets, are among the most powerful non-thermal sources in the Universe. 
In {\it blazars}, having one of the jets pointing at the direction of the Earth,
the Doppler boosted non-thermal output from the jet completely outshines the host galaxy. Their emission
is characterized by strong outbursts (``{\it flares}'') and fast variability, sometimes down to time scales
of minutes.

In the last years, the number of VHE extragalactic sources has been growing steadily, 
the vast majority of the newly discovered sources belonging to this
class of blazars. Several blazar discoveries and results from  multi-wavelength monitoring campaigns
have been presented at the conference.
Among others, the \HESS\ collaboration presented the first \HESS-II\ results, in monoscopic mode, on 
the well know high synchrotron peaked objects PKS~2155-304 and PG~1553+113~\cite{hess-agns-icrc2015}.
The data acquired with CT5 revealed a significant  spectral curvature for both sources with respect 
to a simple power-law spectrum and contributed to close the gap with \LAT.



A few AGNs exhibit particularly interesting characteristics and will be described in the following.

\subsection{IC~310 Lightning from a Black Hole}

\begin{figure}[t!]
\begin{center}
\includegraphics[width=0.49\textwidth]{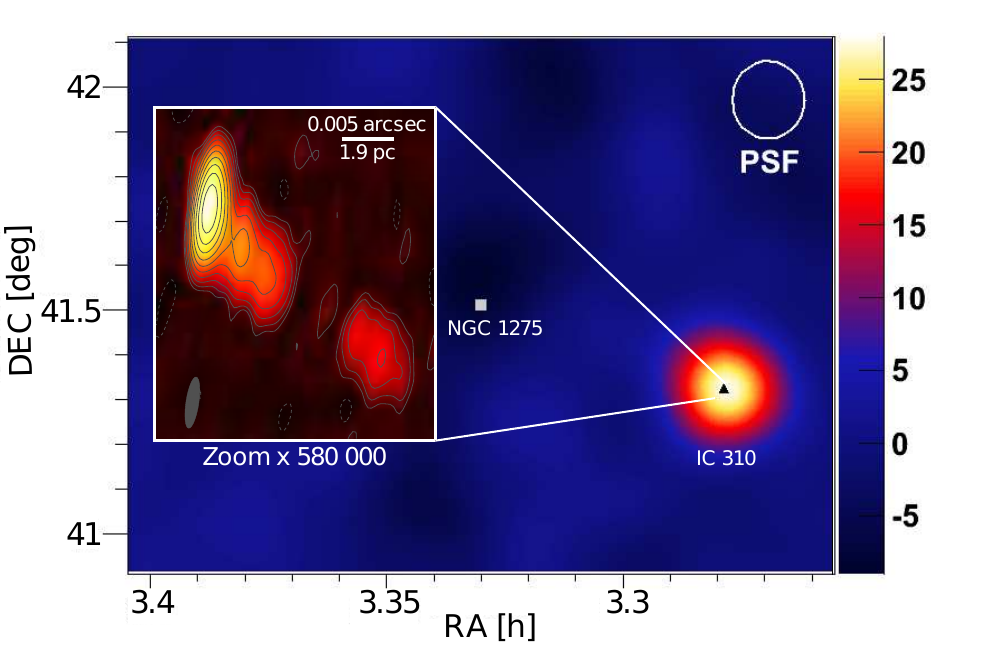}
\includegraphics[width=0.49\textwidth]{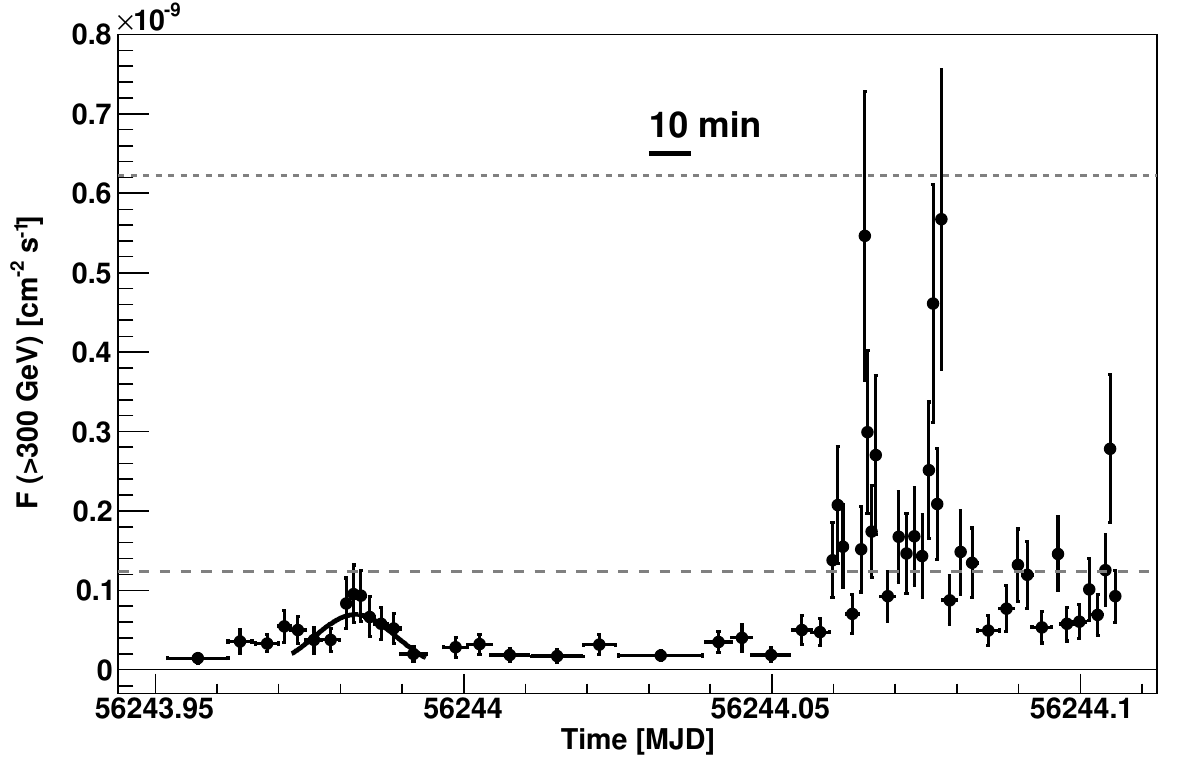}
\end{center}
\caption{\label{fig:IC310}{\bf Left:} Combined image of the significance map measured by MAGIC above $300\U{GeV}$  of the Perseus cluster of galaxies 
and the radio map obtained from EVN observations~\cite{2012A&A...538L...1K}.
{\bf Right:} Light curve above $300\U{GeV}$ of the flare recorded by MAGIC in 2012~\cite{magic-ic310-science,magic-IC310-icrc2015}.}
\end{figure}

IC~310 is a nearby ($z = 0.0189$) radio galaxy, located in the outskirts of the Perseus cluster of Galaxies, 
and hosting a $3\times 10^8 M_\odot$ super-massive black
hole~\cite{magic-ic310-science}. It shows a broad-band non-thermal emission, detected from 
the radio to the VHE $\gamma$ rays~\cite{magic-ic310}. 
IC~310 was originally classified as a head-tail radio galaxy. 
Recent Very Long Baseline Interferometry (VLBI)
established the existence of straight jets, showing the same direction 
from parsec to kiloparsec~\cite{2012A&A...538L...1K},
and revealed a blazar-like inner jet with a missing counter jet (Fig.~\ref{fig:IC310}, left). 
From these observations, IC~310 was classified as a peculiar object in-between blazars and radio galaxies, 
with an angle between the jet-axis and the line-of-sight measured to be $10^\circ \lesssim \theta \lesssim 20^\circ$~\cite{magic-ic310-paper2}.
This was further supported by the measurement of low Doppler factors in the jets with 
radio telescopes~\cite{2012A&A...538L...1K}.

During a multi-wavelength campaign organized by the MAGIC collaboration in 2012, an 
exceptionally bright VHE flare was detected, reaching the flux of the Crab
Nebula above $1\U{TeV}$ and showing very fast variability, down to timescales of a few minutes (Fig.~\ref{fig:IC310}, right).
Due to the non-zero angle of the jet, the variability timescale cannot be significantly lowered by Doppler boosting, 
and therefore allows the size of the emission region to be constrained to be smaller than $20\%$ of the Schwarzschild radius of the central black hole. 
In frequently used AGN models, such as the so-called {\it shock-in-jet} model, such small emission regions would unavoidably 
result in strong absorption of VHE \gray\ emission
by pair creation on the synchrotron radiation (``{\it opacity}'' argument). 

New models have been proposed to circumvent the opacity problem, see~\cite{magic-ic310-science} and references therein, 
including mini-jet structures within the  jets, jet-cloud/star interactions, and magnetospheric models~\cite{2011ApJ...730..123L}. 
The latter, inspired from pulsar models, is able to reproduce the main features of the VHE \gray\ emission. In this model, the black hole
is assumed to spin maximally. A particle-starved magnetosphere is then anchored to the ergosphere (location at which space-time is flowing at the speed of light),
leading to the formation of vacuum gaps where electrons and positrons can be accelerated to ultrahigh energies. Theses develop
electromagnetic cascades, multiplying the number of charge carriers, until the induced current shortcuts the gap.
This analysis however relies almost completely on the measured
jet angle. For a smaller jet angle ($\theta \lesssim 5^\circ$), usual shock-in-jet models would probably be able 
to reproduce the observed features.

\subsection{Quasar half-way to the edge of the observable Universe}

PKS~1441+25 is an flat spectrum radio quasar (FSRQ) located at a red-shift of $z = 0.939$.
First detected in VHEs by MAGIC~\cite{ATEL7416} at a statistical significance of $25\,\sigma$ after a  MeV-GeV flare~\cite{ATEL7402}, 
it was quickly confirmed by VERITAS~\cite{ATEL7433}  (at a statistical significance of $\sim 8\,\sigma$).
The source appeared to be steady at that time, at $\sim 5\%$ of the Crab flux above $80\U{GeV}$ with a very soft spectrum.
In later observations in May 2015, after the end of the MeV-GeV flare, PKS~1441+25 could not be detected at the same level in VHEs.
Being one of the two most distant VHE blazar ever, the source is of particular interest. Detected up to $\sim 250\U{GeV}$, it implies important
constraints on the extragalactic background light (EBL)~\cite{magic-fsrq-icrc2015,veritas-agns-icrc2015}.
With five detected objects only so far, FSRQs currently form only a small fraction of the VHE extragalactic sources.

\subsection{QSO B0218+357: A Gravitationally lensed blazar}

Due to the strong absorption of VHE $\gamma$ rays by pair creation on the EBL,
most of VHE blazars known so far are relatively close-by, with red-shift $z \lesssim 0.5$. Below $100\U{GeV}$,
where the sensitivity of IACTs start somewhat degrading, the observable Universe extends to $z \sim 1$.

QSO~B0218+357 is a distant blazar, likely a flat spectrum radio quasar, located at the red-shift of $z=0.944$~\cite{2012ApJ...744..177L}
and gravitationally lensed by the galaxy B0218+357~G located at the red-shift of $0.68$.
Einstein's rings separated by only $335\U{mas}$ have been detected in radio~\cite{1992AJ....104.1320O}.
Time delays between the two components have been detected in radio (of $10-12\U{days}$)~\cite{1999MNRAS.304..349B}
and recently in HE $\gamma$ rays ($11.46 \pm 0.16\U{days}$), using auto-correlation analysis of the light curve~\cite{fermi-B0218}.

\begin{figure}[ht!]
\begin{center}
\includegraphics[width=0.8\textwidth]{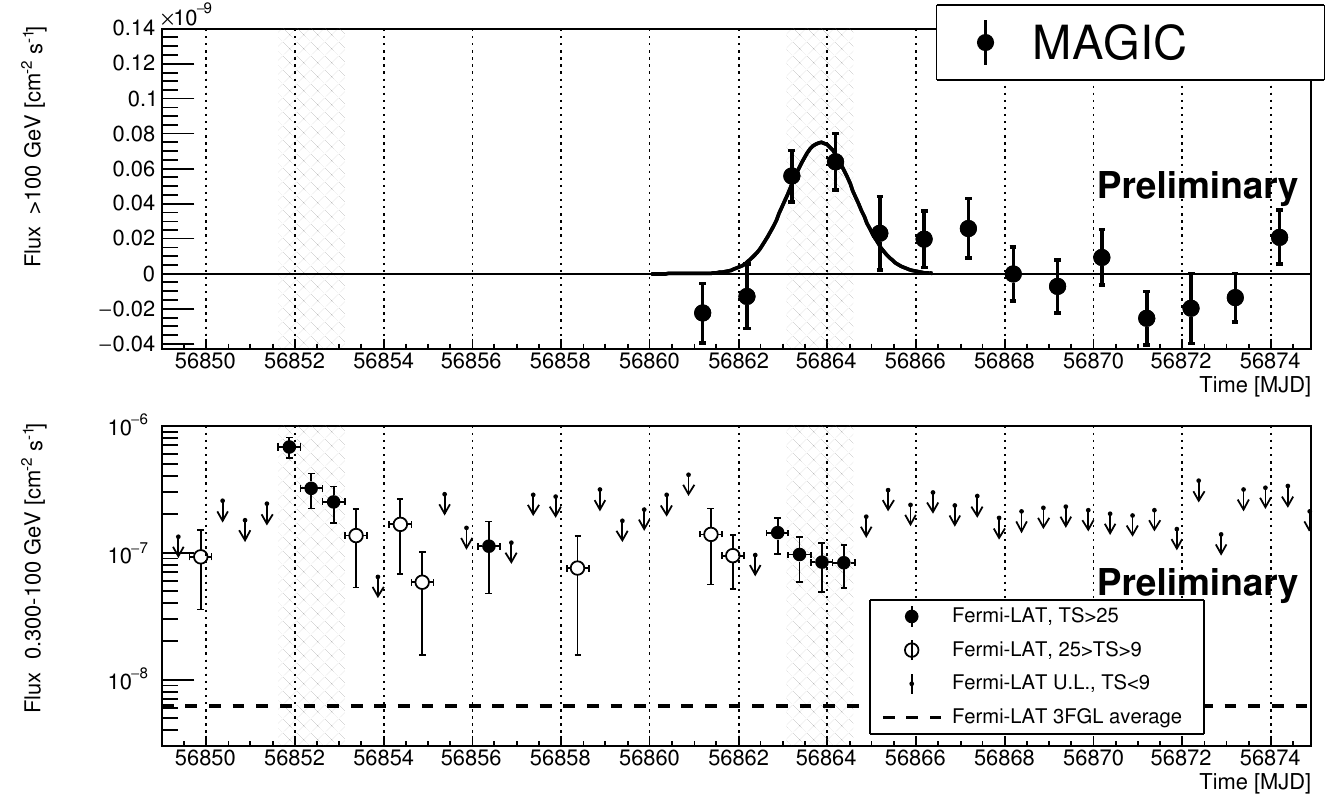}
\end{center}
\caption{\label{fig:MAGIC_GSOB0218}The gravitationally lensed blazar QSO B0218+357. {\bf Top:} MAGIC nightly light curve above $100\U{GeV}$. {\bf Bottom:}
\LAT\ light curve above $0.3\U{GeV}$~\cite{magic-B0218-icrc2015}}
\end{figure}

In July 2014, QSO~B0218+357 entered a flaring state in HE~\cite{ATEL6316}, with an intensity increase by a factor of $\sim 20$. 
Due to full moon period, MAGIC could not detect the original HE flare,
but detected the source at the expected arrival time of the delayed flare~\cite{ATEL6349}, with a statistical significance of $5.7\,\sigma$. 
The corresponding \LAT\ and MAGIC light curves are shown in Fig.~\ref{fig:MAGIC_GSOB0218}.

QSO~B0218+357 is, with PKS~1441+25 one of the two most distant sources (red-shift $z\sim 0.9$) known in VHE $\gamma$ rays,
allowing to derive strong constrains on the EBL.

PKS~1830-211, another gravitationally lensed blazar, was observed with \HESS-II\ in August 2014 following 
an alert issued by the \LAT\ collaboration on a flaring state, to search for a potential VHE counterpart in active 
state with a delay of $\sim 20$ days as detected with \LAT\ in a previous flare from this source~\cite{2015ApJ...799..143A}. 
No detection were achieved based on 8.6 h of data, leaving thus QSO~B0218+357 as the only gravitationally 
lensed blazar detected at the VHE energies.

\subsection{Distant sources \& EBL measurement}

The VHE energy spectrum of distant AGNs is affected by pair creation on the EBL and on the CMB. The imprint depends on the energy and
on the red-shift of the source. Several probes can be used to obtain constrains on the EBL:

\begin{itemize}
\item Single sources at large distance, such as PKS~1441-25 or 1ES~1101+496 can be used to derive upper limits of the EBL density.
\item Measurement of several sources at different distances allow, under some assumption of the intrinsic spectrum, to measure the
EBL density at different distances. This has been done recently~\cite{Biteau2015} using 86 spectra from 38 individual VHE sources,
corresponding to a total of $\sim 270, 000$ $\gamma$ rays. Using a sophisticated statistical treatment, evidence of the imprint of \gray absorption
is found at the $11\, \sigma$ significance level in the energy spectra of VHE blazars.
The reconstructed EBL spectrum is in good agreement with estimates based on
galaxy counts (below direct measurements), and leaves room for contributions from
e.g. intra-halo light. Several results presented at the conference supported this conclusion~\cite{veritas-agns-icrc2015,magic-highlights-icrc2015,magic-fsrq-icrc2015}.
\end{itemize}

\begin{figure}[ht!]
\begin{center}
\includegraphics[width=0.8\textwidth]{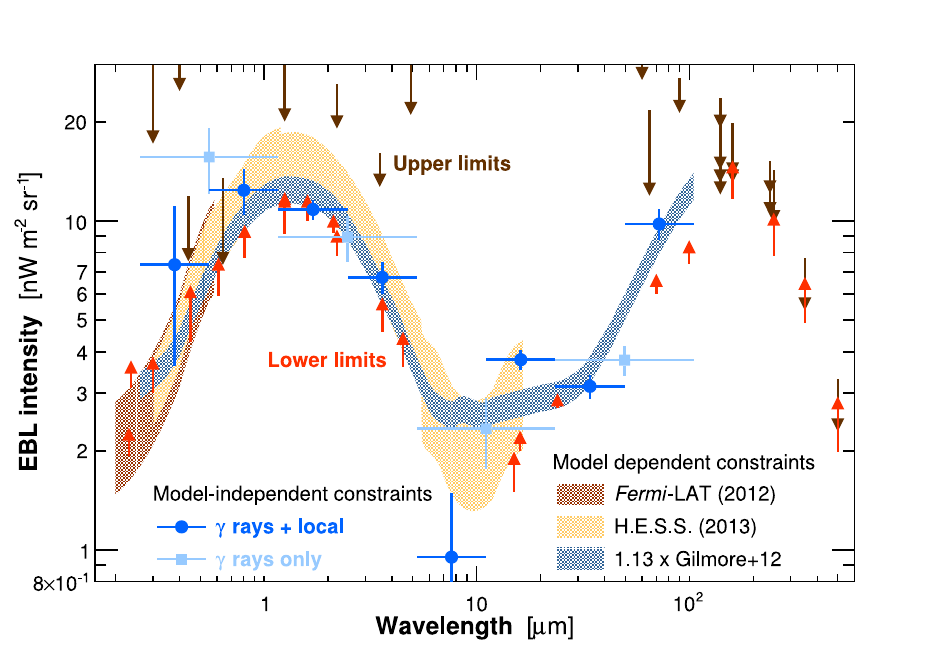}
\end{center}
\caption{\label{fig:EBLBiteau}Local EBL measurement using a large sample of VHE blazars~\cite{Biteau2015}.}
\end{figure}

\section*{Conclusions and perspectives}

\begin{figure}[htb!]
\begin{center}
\includegraphics[width=0.9\textwidth]{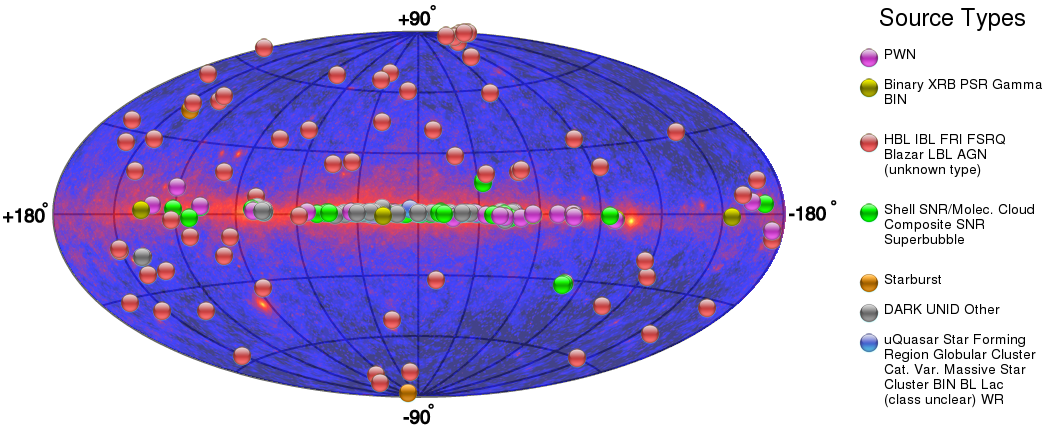}
\end{center}
\caption{\label{fig:TevCat}Current catalogue of VHE sources. From \protect\href{http://tevcat.uchicago.edu/}{http://tevcat.uchicago.edu/}.}
\end{figure}

\begin{figure}[htb!]
\begin{center}
\includegraphics[width=0.9\textwidth]{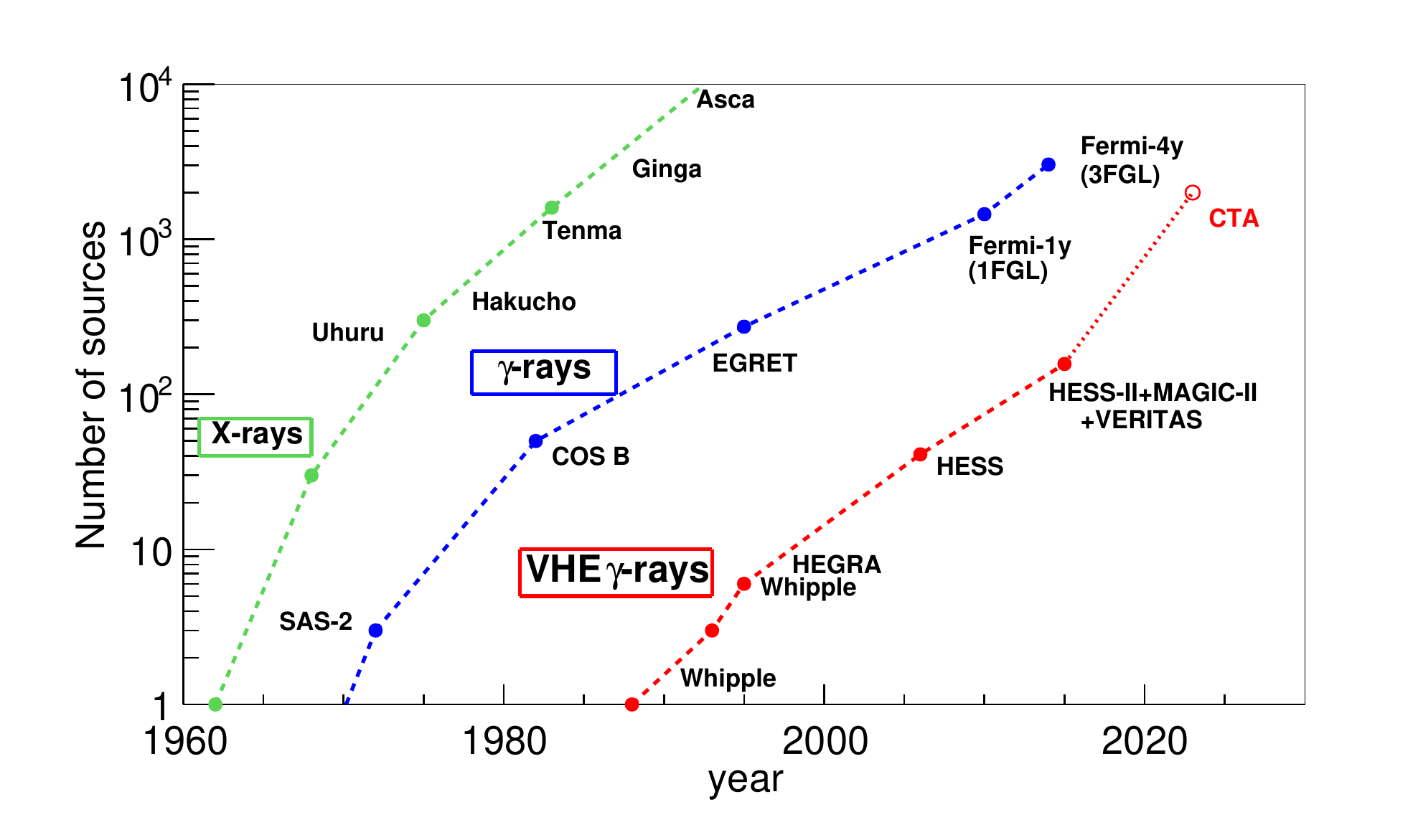}
\end{center}
\caption{\label{fig:Kifune}The so-called Kifune plot showing evolution of number of sources
at different wavelengths as a function of time and different instruments~\cite{CRAS_IACTS}.}
\end{figure}

Nearly 10 years after the coming in of the third generation instruments, the field of ground based \gray\ astronomy is still very active.
Numerous technical developments, together with elaborated analysis techniques, have resulted in major improvements of the sensitivity of
the instruments~(Fig.\ref{fig:SensVsTime}). This translated almost directly into the number of detected sources at VHEs, as shown 
in the current VHE catalogue (Fig.~\ref{fig:TevCat}).
The number of discoveries as function of time is shown in Fig.~\ref{fig:Kifune}.
Similarly to the evolution of the X-ray domain in the 1960's, there is almost an exponential rise of the number of sources, without any saturation so far.
For many researchers, the recent success of VHE astronomy comes as a surprise, as in the 1980's only a handful
of them  believed there was any detectable \gray source above $100\U{GeV}$. 

The field is however experiencing a phase transition, as most of the brightest sources are probably already detected. Large surveys
and deep exposures of specific sources are now defining the general trend.
At the same time, large multi-wavelength observation
campaigns, involving many different observatories operating across the electromagnetic spectrum, proved to be the most effective way to unveil
the nature of the very high energy variable sources. With the advent of the Cerenkov Telescope Array (CTA), this will most likely 
be more and more the case. 
A re-organization of the collaborations, similar to that of optical observatories, is unavoidable.
Open calls for  observation proposals and data rights either partially or completely public 
will probably become the next paradigm.

Despite the numerous successes of VHE ground based astronomy, some important results are still awaited:

\begin{itemize}
\item Dark matter, predicted in many theories to annihilate in pairs of high energy $\gamma$ grays, has not been detected so far. Recent limits
from Dwarf Galaxies are becoming very constraining.
\item Despite huge efforts to improve the relocation speed and the energy threshold of the telescopes and to minimize the dead time 
between consecutive observations, and despite evidence from \LAT\ of emission up to $\sim 100\U{GeV}$~\cite{fermi-grb130427A},
still Gamma Ray Bursts (GRBs) remain undetected at VHEs~\cite{magic-grb-icrc2015}.
\item Definite proof of acceleration of hadrons, though the detection of very high energy neutrinos from astrophysical sources,
are still missing.
\item Sources of Ultra High Energy cosmic rays, above $10^{15}\U{eV}$, are still unidentified. SNRs are commonly thought
to be able to accelerate cosmic rays up to the knee, however no {\it Pevatron} has been firmly identified yet.
\end{itemize}

With an improvement by a factor of $\sim 10$ in sensitivity, and with a significantly better angular resolution, 
CTA~\cite{CTA:2013a} will be the next breakthrough in VHE ground based \gray\ astronomy.
Construction phase should begin in 2016 with the goal to operate the full array by 2021. Exciting discoveries
are still ahead of us.

\section*{Acknowledgements}

The author is very grateful to the \HESS, MAGIC and VERITAS collaborations for providing their results. He would like
to thank in particular Reshmi Mukherjee and Razmik Mirzoyan for very interesting and fruitful discussions.



\clearpage

\bibliographystyle{PoS} 
\addcontentsline{toc}{part}{Bibliography}
\bibliography{iact-techniques,hess,cangaroo,bibtex_db,whipple,veritas,magic,sampling,algorithms,shower,milagro,hawc,cras,binaries,snr,pulsars,hegra,fermi,egret,lmc,agns,agile} 

\end{document}